\documentclass[useAMS]{mn2e}

\title[PKS2250-41: a case study for triggering]{PKS2250-41: a case study for triggering\thanks{Based on observations collected at the European
    Southern Observatory, Chile (programs 071.B-0320(A), 074.B-0310(A), 075.B-0820(A)
    and 078.B-0500(A))}}
\author[K.\,J. Inskip et al ]{K. J. Inskip$^{1}$\thanks{E-mail:
K.Inskip@shef.ac.uk}, M. Villar-Mart\'{i}n$^{2}$, C. N. Tadhunter$^{1}$,
  R. Morganti$^{3}$, J. Holt$^{1,4}$ and \newauthor D. Dicken$^{1}$ \\
$^{1}$ Department of Physics \& Astronomy, University of Sheffield, Sheffield S3 7RH\\
$^{2}$ Instituto de Astrof\'{i}sica de Andaluc\'{i}a (CSIC),
  Aptd0. 3004, 18080 Granada, Spain\\
$^{3}$ Netherlands Foundation for Research in Astronomy, Postbus 2,
  7990 AA Dwingeloo, The Netherlands\\
$^{4}$ Sterrewacht Leiden, Postbus 9513, 2300 RA Leiden, The Netherlands}
\begin{document}

\date{}

\pagerange{\pageref{firstpage}--\pageref{lastpage}} \pubyear{2007}

\maketitle

\label{firstpage}

\begin{abstract}We present the results of a multiwavelength study of
the $z = 0.31$ radio source PKS2250-41.   Integral field unit and long-slit
spectroscopy obtained using VIMOS and FORS1 on the VLT, and archival
HST optical imaging observations are
used to study the morphology, kinematics and ionisation state of the
extended emission line region (EELR) surrounding this source, and also
a companion galaxy at a similar redshift.  Near-infrared
imaging observations obtained using the NTT are used to analyse the
underlying galaxy morphologies. The EELR displays a complex variety of
different gas kinematics and ionization states, consistent with a mixture
of radio source shocks and AGN photoionization.

The radio galaxy is likely to lie within a group environment, and is
plausibly undergoing interactions with one or more other objects.
The
disk-like galaxy to the northeast of the radio source lies at a
similar redshift to the radio galaxy itself, and has its major axis
position angle aligned with the filamentary continuum and line
emission extending outwards from the radio galaxy.  This filamentary
structure is most plausibly interpreted as a tidal structure
associated with an interaction involving the radio source host galaxy
and the aligned companion galaxy to the north-east; this encounter may
have potentially triggered the current epoch of radio source
activity.  Overall, PKS2250-41 displays some of the best evidence that
radio source activity can be triggered in this manner.

While the environment and recent interactions of a radio galaxy can have
some bearing on its subsequent evolution, our data also highlight
the varied means by which the radio source can effect changes in
adjacent objects.  Our
IFU and long-slit spectroscopy confirm the presence of radio source
shocks within the western radio lobe, and, together with our continuum
observations, add further weight to the 
presence of a faint continuum source coincident with the secondary
hotspot in the western radio lobe.  On the basis of our
multiwavelength observations of this object, we suggest that the radio
source has indeed triggered recent star formation within this faint companion.

\end{abstract}


\section{Introduction}

The presence of a powerful radio source is known to have profound
implications for the properties of both the host galaxy and the
surrounding intergalactic medium (IGM).  While the host galaxies are usually massive
ellipticals with predominantly old stellar populations 
formed at $z \sim 5-20$ (e.g. Inskip et al 2002a), increasing evidence is accumulating for more
recent star formation in many sources (e.g. Wills et al 2007, Holt et
al 2007 and references therein).
Extended emission line regions (EELRs) are frequently observed around
radio galaxies (McCarthy et al 1987), and their observed
properties (size, luminosity, kinematics and ionisation state) are
known to depend strongly on those of the radio source (e.g. Best et al
2000, Inskip et al 2002b,c, Moy \& Rocca-Volmerange 2002, Humphrey et
al 2006).  However, the
origin of the emitting material is less well understood.
Overall, disentangling cause from effect within these complex systems
is far from straightforward; which 
properties are genuine symptoms of the radio source activity, and which
can be tied to the underlying AGN triggering mechanism? 

Integral field spectroscopy (IFS) provides a powerful tool for addressing
questions such as these. The additional spatial data provided by IFS 
observations is a major advantage, as it allows the EELR properties
(physical conditions, gas dynamics, ionisation state) to be
efficiently studied and quantified as a function of  position relative
to the radio source, rather than just along the radio axis, as is
commonly the case for most long-slit spectroscopic studies.  
In this paper, we consider the case of the $z \sim 0.31$ radio galaxy
PKS2250-41, previous observations of which have shown it to be surrounded by
one of the most impressive EELRs identified at intermediate
redshifts (Tadhunter et al 1994; Clark et al 1997; Villar-Mart\'{i}n
et al 1999).  We present the results of a multi-wavelength study of this
source, combining optical IFS spectroscopy, optical
long-slit spectroscopy, near-IR imaging and optical continuum and
emission line imaging.

The details of the source, our
observations and data reduction are described in section 2, and the results
presented in section 3, including an analysis of potential companion
objects at a similar redshift.  In section 4, we 
discuss the implications of our results, and we present our
conclusions in section 5. 
Throughout this paper, we assume
cosmological parameters of $\Omega_0 = 0.3$, $\Omega_{\Lambda} =
0.7$ and $H_0 = 70 \rm{km\,s^{-1}\,Mpc^{-1}}$, which result in  an
angular scale of $\sim 4.55$kpc/arcsec at $z \sim 0.31$. 



\section{Source details, observations and data reduction}

\subsection{Source details}

\begin{figure}
\vspace{4.85 in}
\begin{center}
\includegraphics{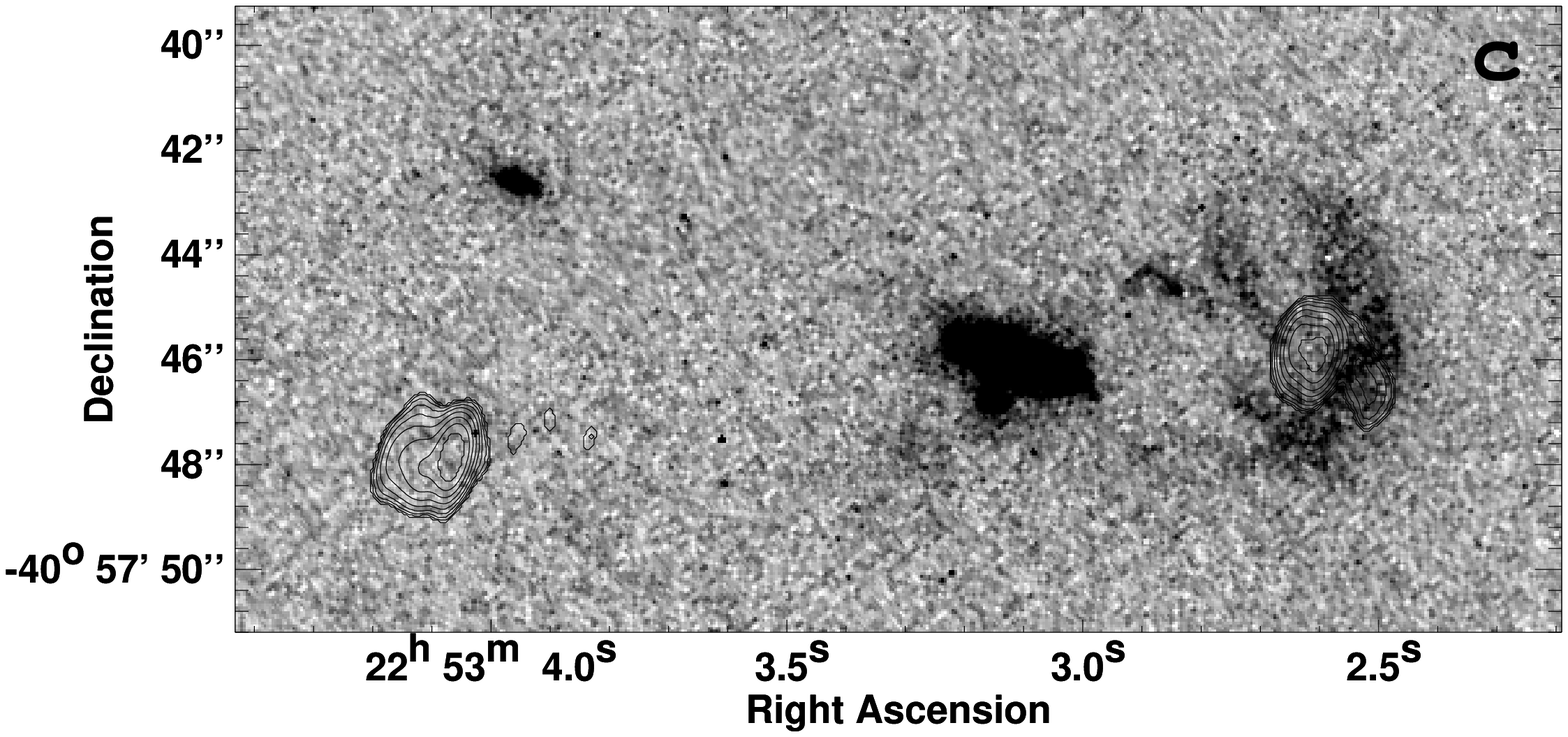}
\includegraphics{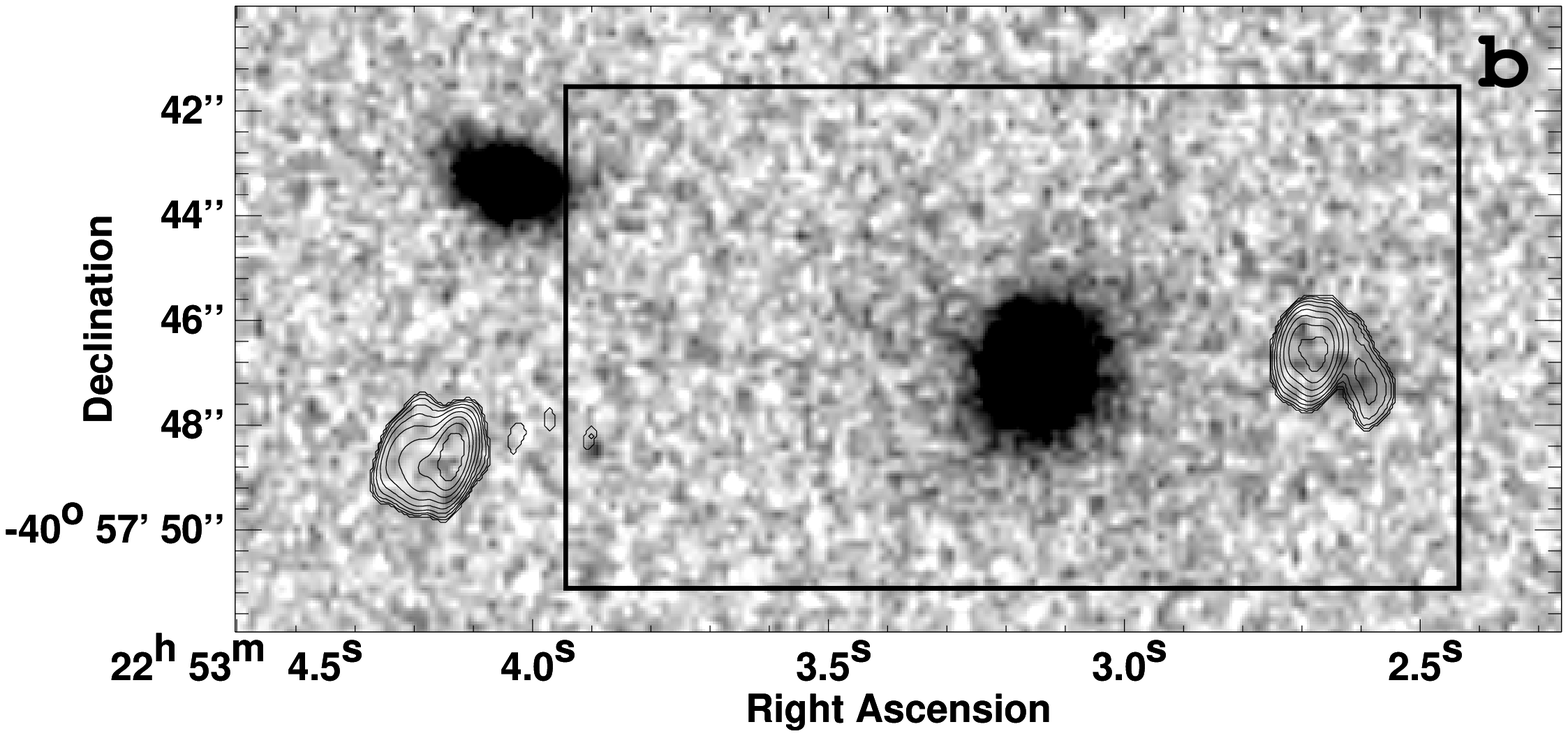}
\includegraphics{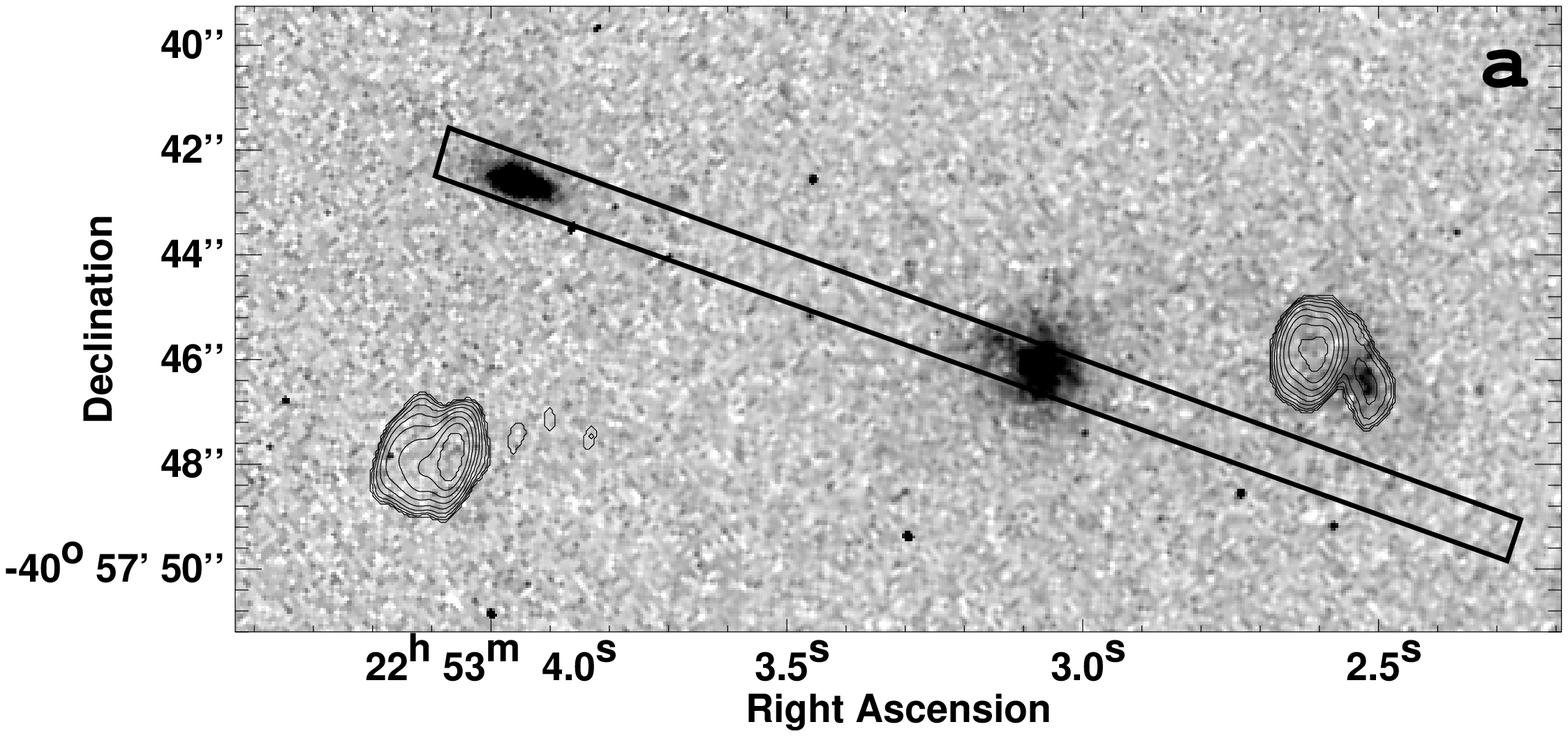}
\end{center}
\caption{
(a - top) Optical WFPC2 F547M image of PKS2250-41,
overlaid with 15GHz VLA radio contours, also illustrating the location of the
PA 69$^{\circ}$ $1.3^{\prime\prime}$ FORS1 slit. (b - centre) Infrared
SOFI Ks image of PKS2250-41, overlaid with 15GHz VLA radio contours, also illustrating the region of the VIMOS IFU
field of view studied in detail in this paper. (c - bottom) WFPC2
[O\textsc{iii}]5007\AA\ emission line image of  PKS2250-41,
overlaid with 15GHz VLA radio contours. The WFPC2 and radio
data are taken from T05. The major features to note in frame (c) are
the extensive emission line arc to the west, coincident with the radio
emission in the western lobe, and a bright linear structure running
NE-SW across the host galaxy, aligned towards the companion galaxy to
the NE.  
\label{Fig: radio}}
\end{figure}

PKS2250-41 is an FRII radio source hosted by an elliptical galaxy
lying at a redshift of $z = 0.308$, and a member of the large sample of southern 2Jy
sources selected for study by Morganti, Killeen and Tadhunter (1993)
and Tadhunter et al (1993).  Fig.~\ref{Fig: radio} displays the most
recent optical and infrared imaging of this source, overlaid with VLA
radio contours.

The EELR surrounding PKS2250-41 is particularly striking, and has been
well-studied in the past. Imaging, spectroscopic and polarimetric
observations of this source were obtained in 1993, and the results
presented in a number of papers (Tadhunter et al 1994, Shaw et al 1995 and Clark et al
1997).   The major feature of this source is the bright emission
line arc lying approximately 5.5 arcsec west of the host galaxy,
circumscribing the radio lobe, with lower surface brightness line
emission lying to the east of the host galaxy.  The varied gas kinematics and ionization
states observed in the western arc provide some of the best evidence
in any active galaxy for the presence of shocked gas with a
photoionized precursor region, plausibly caused by jet-cloud
interactions.  Emission line density and temperature diagnostics imply high 
pressures in the western radio lobe, suggesting pressure equilibrium
within this region. After subtraction of the nebular continuum
emission from this region, Dickson et al (1995) found that the
remaining continuum is relatively blue in colour, while Shaw et al
(1995) showed that it is also unpolarised. 
Clark et al (1997) hypothesise that the western
arc originates from a direct collision between the radio jet and a
companion object.  They also suggest that the galaxy lying at a
distance of approximately 10.5 arcsec/50 kpc to the northeast of PKS2250-41
lies at a similar redshift to the radio source and that the two
galaxies may be undergoing some level of interaction.  It should
be noted that their
data were of lower signal-to-noise, and were not definitive
with regard to the redshift of the companion.

On the basis of the clear evidence for jet-cloud interactions
displayed by this source, Villar-Mart\'in et al (1999; hereafter VM99)
carried out a further, higher resolution spectroscopic study utilising four different
slit positions, three aligned on, above
and below the host galaxy centroid and parallel to the radio axis, and one perpendicular to the radio
source axis, aligned with the western emission-line arc.  Two
different kinematic components were identified within the gas, and the
other emission line properties (temperature, ionization state,
ionization mechanism) clarified, adding strong support to the
picture of cooling gas behind a shock front driven by the radio jets,
as well as a photoionized
precursor region. 

\begin{table}
\caption{Details of the observational data used in this paper.  The
  listed seeing measurements are given for the wavelength
  of the observations in question.
}
\begin{center} 
\begin{tabular}{ccccc}
Date & Instrument & Time (s) & Seeing & Notes \\\hline
2003/07/24 & FORS1  & 6720 & 0.4-0.7$^{\prime \prime}$& polarimetry\\
2004/10/19 & VIMOS & 13560 &1.1-1.5$^{\prime\prime}$& cirrus\\
2004/11/19 & VIMOS  &9040&1.1-1.5$^{\prime\prime}$& photometric \\
2005/06/07 & VIMOS  &17600&0.7-1.1$^{\prime\prime}$ & photometric\\
2005/06/09 & VIMOS  &8800&0.5$^{\prime\prime}$ & photometric\\
2006/11/14 & SOFI & 3000 & 0.8$^{\prime\prime}$& photometric\\
\end{tabular}
\end{center} 
\end{table}

The subsequent emission-line and continuum imaging of Tilak et al (2005; hereafter
T05) is in keeping with this picture.  The different
ionization states within the arc are spatially and spectrally resolved, suggesting a
cooling-region behind the radio source bow-shock.  Continuum emission
is potentially observed at the location of the secondary hotspot, and
a continuum spur extending northeastwards from the host galaxy is
identified, aligned with the brightest emission line features in that region.

The observations of PKS2250-41 presented in this paper  include optical
long-slit and integral field unit spectroscopy, and infrared imaging.
The primary goals of these observations are to deepen our
understanding of the EELR surrounding this source with regards to its
origin, including both the 
western emission-line arc and the filamentary spur extending towards
the northeast, and additionally to investigate the nature of the
possible companion galaxies.   The observational data presented in
this paper is outlined in full in the remainder of this section, with
brief details provided in Table 1.

\subsection{FORS1 spectroscopic observations}

Long-slit spectropolarimetry observations of PKS2250-41 were obtained on 2003
July 24 using the  Focal Reducer/low dispersion Spectrograph (FORS1;
Appenzeller et al 1998) as part of the European Southern
Observatory (ESO) observing programme 071.B-0320(A).  FORS1 is mounted
on the Nasmyth focus of the UT2 Kueyen unit of the Very Large Telescope (VLT). 
The instrument was used with the GG435 filter and GRIS\_600R grism,
giving a spectral resolution of 1.07\AA/pixel, an instrumental
resolution of $5.9 \pm 0.1$\AA, a plate scale
of 0.2 arcsec/pixel and an overall observed-frame useful spectral coverage of $\sim
5520-7410$\AA.  The slit width during the observations was 1.3 arcsec.  Sixteen separate 420s observations were made, in 
excellent seeing conditions of 0.4-0.7$^{\prime\prime}$.  
The position angle of the FORS1 slit was $69^\circ$, and is illustrated in Fig.~\ref{Fig: radio},
overlaid on a ground-based emission line plus continuum image of
PKS2250-41 (Clark et al 1997).  In addition to including flux from the
host galaxy and a substantial portion of the EELR, the slit PA was
chosen so as to also include flux from the possible companion galaxy
10.5 arcsec to
the northeast.  

Standard packages within the NOAO \textsc{iraf} reduction software
were used to reduce the raw data. Corrections were made for
bias subtraction, and the data then flat-fielded using internal
calibration lamp observations.  The flat field images were made using
the same instrumental set-up as the observations for each
source.  The 2-d spectra were corrected for geometric distortion,  and then wavelength
calibrated.  The frames obtained in the 'o' and 'e' polarisation
states were spatially aligned, combined, and cosmic rays
removed.  Polarisation results will be presented elsewhere
(Tadhunter et al, in prep); in this
paper we concentrate on the gas kinematics and ionization state.
Flux calibration was carried out using observations of the
spectrophotometric flux standard star LTT377, and the data were also
corrected for atmospheric extinction.  The data were corrected for Galactic extinction using 
$E(B-V)$ values for the Milky Way from the NASA Extragalactic
Database (Schlegel et al 1998), and the empirical extinction function
of Cardelli et al (1989).

\begin{figure}
\vspace{1.3 in}
\begin{center}
\includegraphics{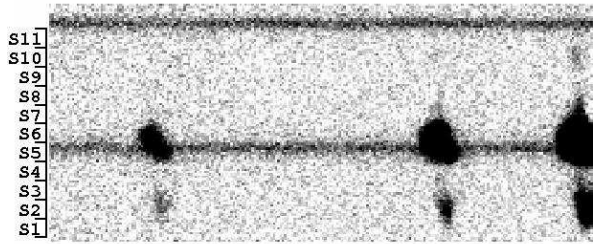}
\end{center}
\caption{Section from the 2-d FORS1 spectrum of PKS2250-41 at
  PA$-69^{\circ}$, covering the rest-frame wavelength range
  $4860-5080$\AA. The spectrum of the companion galaxy to the north
  east lies to the top of the image, and the slit intersects the
  western emission-line arc towards the bottom of the image. The
  emission lines visible on this image (H$\beta$,
  [O\textsc{iii}]4959\AA, [O\textsc{iii}]5007\AA) illustrate the
  complexity and extent of the EELR surrounding this source.  The
  locations of our extracted apertures across the EELR are illustrated on the left
  hand side of this figure. 
\label{2dFORS_spec}}
\end{figure}

Fig.~\ref{2dFORS_spec} displays a limited section of
the two-dimensional FORS1 spectrum (including the H$\beta$ and
[O\textsc{iii}]4959+5007\AA\ emission lines); the companion object continuum is
clearly visible at the top of the image.  It is also noteworthy that
we detect faint very extended line emission between the two galaxies
(see also Clark et al 1997).  


\subsection{VIMOS-IFU spectroscopic observations}

Integral field unit (IFU) spectroscopic observations were carried out
using the Visible Multiobject Spectrograph (VIMOS; Le F\'{e}vre et al
2003, Scodeggio et al 2005), on 2004 October 19, 2004 November 19,
2005 June 7 and 2005 June 9, as part of the ESO observing programmes 074.B-0310(A) and 075.B-0820(A).
VIMOS is mounted on the Nasmyth focus of the UT3 Melipal unit of the
VLT.  These data were obtained using the medium
resolution (MR-orange) grism; using this configuration, the IFU
consists of 1600 microlenses coupled to 0.67-arcsec diameter fibres,
and covers a total sky area of $27 \times 27$ arcsec$^2$.  The total
useful wavelength range is $\sim 4500-9000$\AA, and the spectral
resolution (as determined from unblended skylines) was 
$7 \pm 0.5$\AA. Two separate pointings were used, separated in RA and 
declination by $\sim 2$ arcsec and $\sim 11$ arcsec respectively.
The total integration time was $\sim 49000$s; full details of the
observations and observing conditions are given in table 1.  

Data reduction was carried out using the VIMOS Interactive Pipeline
and Graphical Interface (\textsc{vipgi}) software package (Scodeggio
et al 2005, Zanichelli et al 2005).  Our exact method includes some
variations from the standard \textsc{vipgi} procedures particularly
during the sky-subtraction phase, and is
detailed in full in Inskip et al (2007).

In this paper we concentrate on a limited region of the IFU data-cube,
as illustrated by the box marked on Fig.~\ref{Fig: radio}.

\subsection{SOFI Ks-band imaging}

Infrared Ks-band imaging of PKS2250-41 was carried out on 2006
November 14 using the Son of ISAAC (SOFI; Moorwood, Cuby \& Lidman 1998) instrument
on the ESO 3.5-m New Technology Telescope (NTT).  The instrument was
used in the small field mode, which results in a plate scale of 0.144
arcsec per pixel.  Fifty 1-min exposures were obtained, with each
observation subject to a random offset within a 40 arcsec diameter box.
The data were dark subtracted and corrected for SOFI's interquadrant row cross talk effect
using an adapted version of the SOFI crosstalk.cl IRAF script.
Flat-fielding of the data was carried out using the following process:
all target frames were combined, median filtered and normalised to a
mean pixel value of 1.0 to create a first-pass flat-field image, which
was then applied to each frame.  Bright objects on the flat-fielded
images were then masked out, and the process repeated with the masked
frames, allowing the data to be cleanly flat fielded without any
contamination from stars or galaxies. The subsequent data reduction
uses well-established techniques: the flat-fielded data were
sky-subtracted and combined using the IRAF package DIMSUM, creating a
final mosaiced image of approximately $180 \times 180$ arcsec$^2$.
Flux calibration used observations of NICMOS Photometric Standard
stars (Persson et al 1998), giving a photometric zero-point magnitude
of $K_S = 22.347 \pm 0.010$.  The data were corrected for galactic
extinction using  $E(B-V)$ values for the Milky Way from the NASA
Extragalactic Database, and the parametrized galactic extinction law
of Howarth (1983). 

\subsection{WFPC2 emission line imaging}

In this paper, we will also make use of the previously published WFPC2 medium
band and narrow band imaging of T05.  The emission line
imaging data were obtained using the HST WFPC2 linear ramp filters,
with the target galaxy placed on the region of the chip sensitive to
the observed-frame wavelength of
either the [O\textsc{iii}]5007\AA\ or [O\textsc{ii}]3727\AA\ emission
lines. An intermediate-band continuum image was obtained using the
F547M filter, to allow for continuum subtraction from the
emission-line images, and examination of the continuum structures.  We
refer the reader to T05 for full details of the observations and the data
reduction procedures used.

\section{Results}
\subsection{Results from the FORS1 spectrum}

The 2-d profile of the major emission lines (Fig.~\ref{2dFORS_spec},
see also Fig.~\ref{2250spec}) is
very asymmetric, reflecting the complex kinematics of the EELR.
Relatively broad (up to $\sim 400 \rm km\,s^{-1}$) line emission is observed 
close to the host galaxy, with a clear linear velocity shift. The
line emission to the northeast is blueshifted compared to that lying to the
southwest of the host galaxy continuum centroid.  Between the host galaxy and
the companion, faint, narrower line emission is observed throughout
the intervening space; this emission appears to have a uniform
velocity shift at all spatial offsets.  To the southwest of the host
galaxy, we see the expected features of the emission line arc: line
widths broadening to a maximum at the position of the peak line
intensity. 

\begin{figure}
\vspace{7.05 in}
\begin{center}
\includegraphics{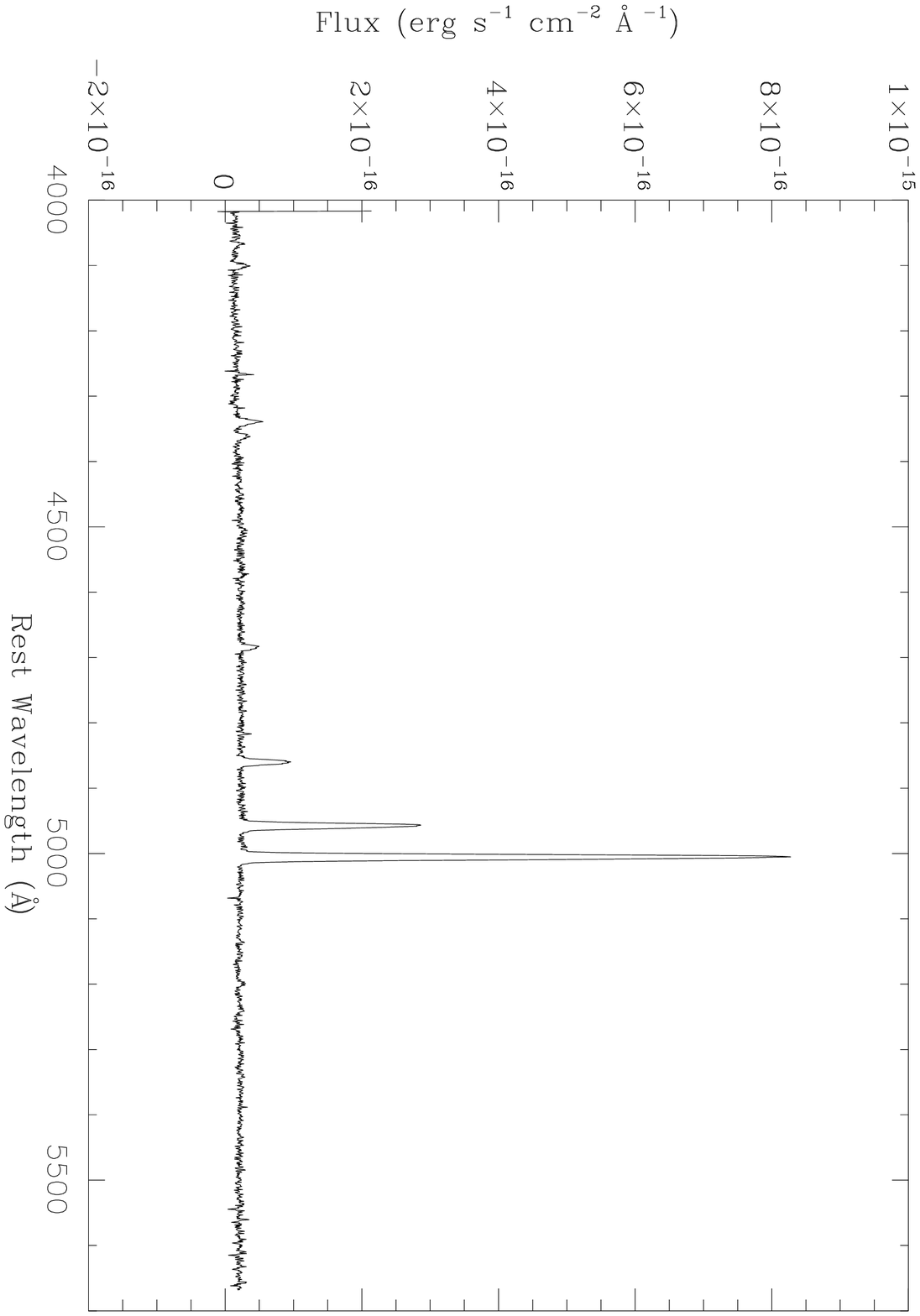}
\includegraphics{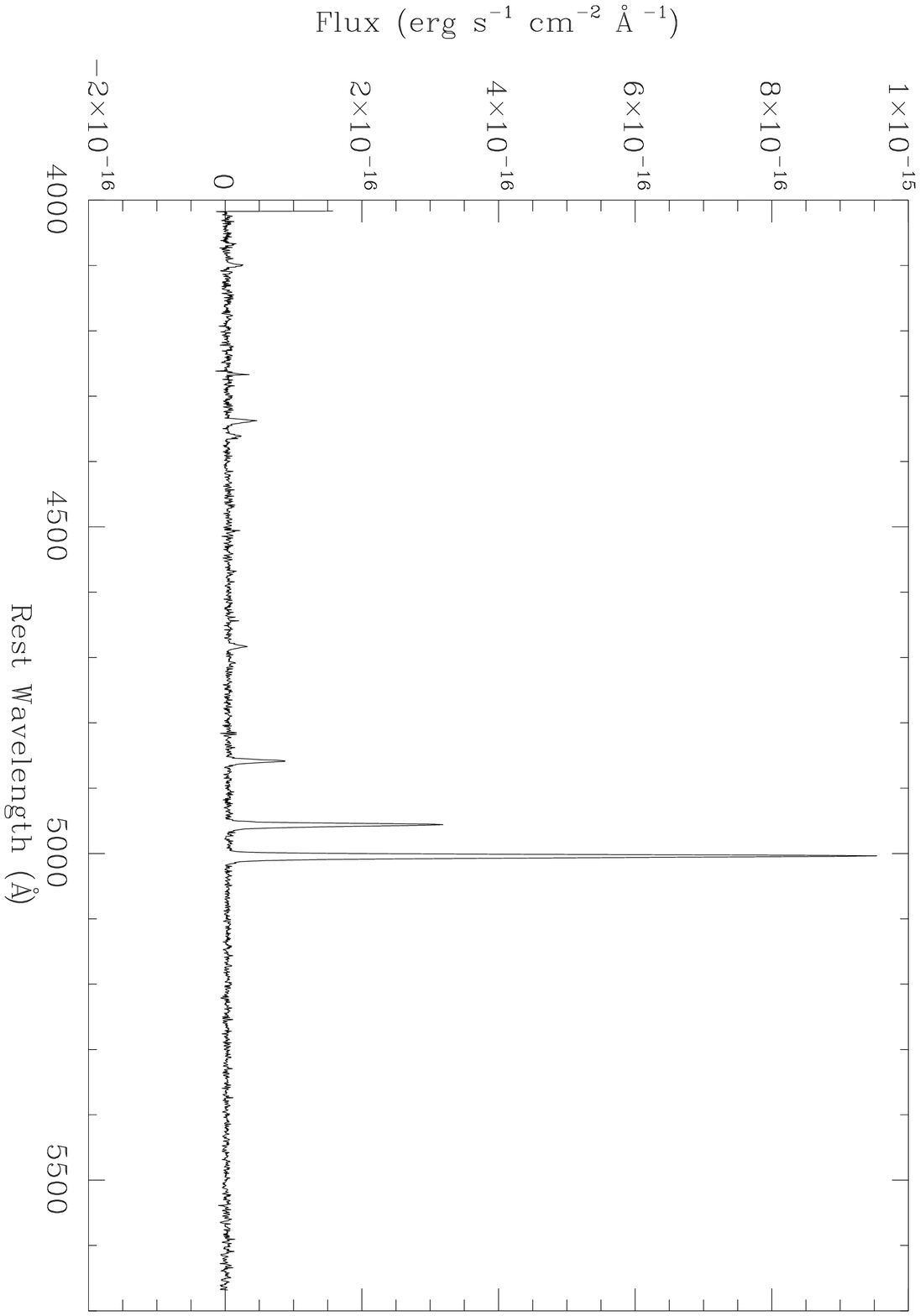}
\includegraphics{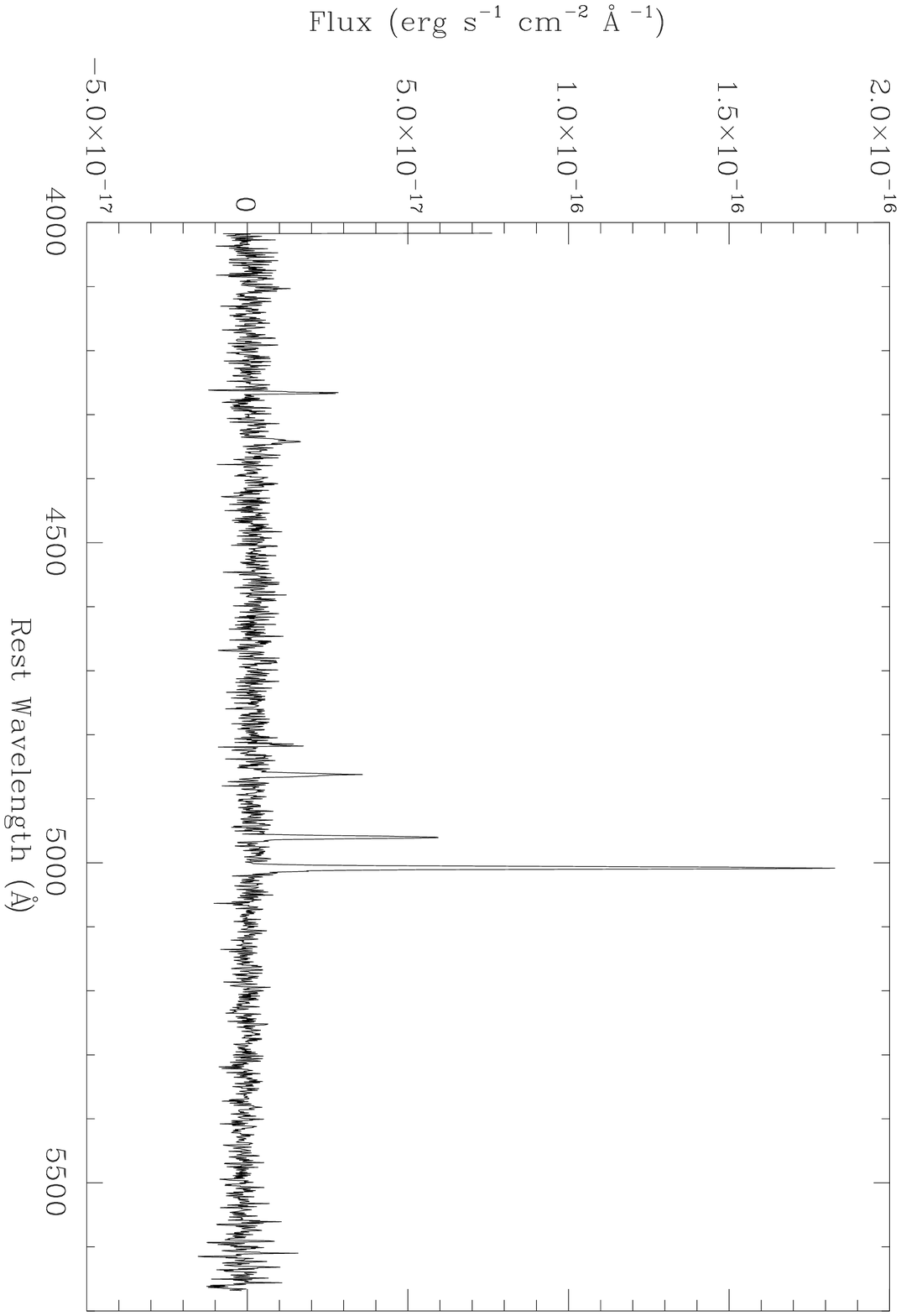}
\end{center}
\caption{Examples of the FORS1 spectra of PKS2250-41, extracted from
  three different $1.6^{\prime\prime}$ apertures: (a - top) centred on the continuum
  centroid, (b - centre) the narrower photoionized filamentary emission
  northeast of the nucleus, and (c - bottom) material in the shocked emission-line arc.
\label{2250_spec}}
\end{figure}

\begin{figure*}
\vspace{4.85 in}
\begin{center}
\includegraphics{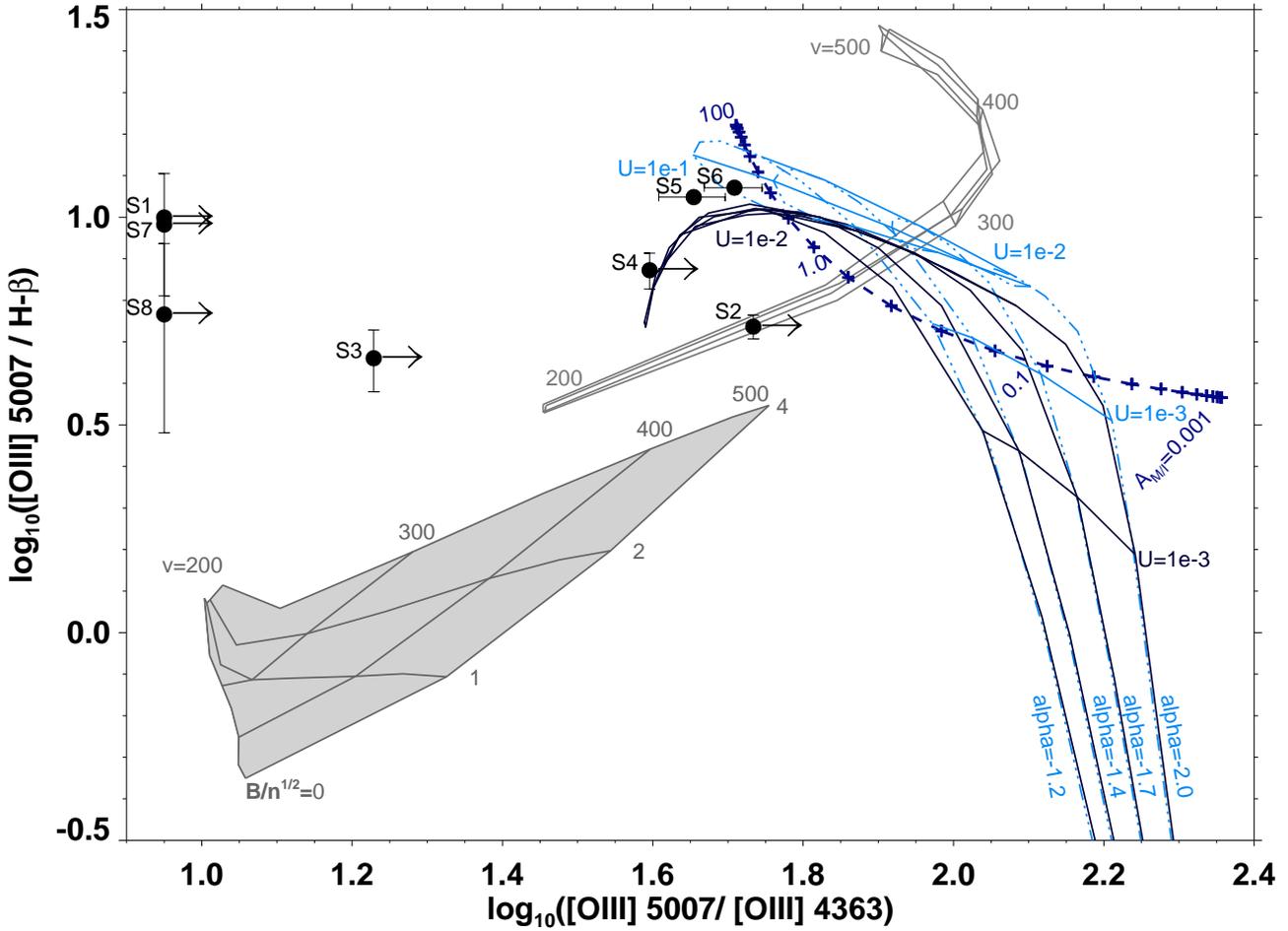}
\end{center}
\caption{Emission line diagnostic plot for the PA-69 FORS1 spectrum of
PKS2250-41, using the line ratios
[O\textsc{iii}]5007/[O\textsc{iii}]4363 and [O\textsc{iii}]5007/H$\beta$.  
Data points represent spectra extracted in 1.6$^{\prime\prime}$ steps
along the slit (see Table 2 and Fig.~\ref{2dFORS_spec} for details).  Lower limits are
given for the [O\textsc{iii}]5007/[O\textsc{iii}]4363 line ratio in
the cases where the [O\textsc{iii}]4363\AA\ emission line is
undetected.  The model tracks
include: (1) shock ionization (Dopita \& Sutherland 1996), with and
without a precursor photoionization region (unshaded and shaded
respectively), (2) the matter-bounded photoionization track (dashed 
dark blue track, Binette, Wilson \& Storchi-Bergmann 1996), and (3)
simple AGN photoionization tracks (light blue tracks), including the
effects of dust (solid black tracks; Groves, Dopita \& Sutherland
2004a,b). The photoionization models are for n=1000 cm$^{-3}$, solar
metallicity and various values of the ionization index $\alpha$ (from
$1.2 < \alpha < 2.0$). 
\label{linediag}}
\end{figure*}

We have extracted a number of one-dimensional spectra from this 2-d
spectrum, in consecutive $1.6^{\prime\prime}$ apertures (Fig.~\ref{2dFORS_spec}).
Fig.~\ref{2250_spec} presents the extracted 1-d spectrum across the
nucleus of the host galaxy.  
%
%
The resulting emission line ratios (see appendix) have been used to
investigate the 
ionization state of the gas, via the temperature sensitive
[O\textsc{iii}]5007/[O\textsc{iii}]4363
vs. [O\textsc{iii}]5007/H$\beta$ diagnostic diagram
(Fig.~\ref{linediag}). Where the [O\textsc{iii}]4363\AA\ was too faint
to be accurately measured, we use the 1-sigma upper limits for the line
flux to determine a lower limit for the
[O\textsc{iii}]5007/[O\textsc{iii}]4363 emission line ratio.

Close to the host galaxy centroid (extracted spectra S4, S5 and S6)
the emission line ratios are consistent with AGN photoionization.
The [O\textsc{iii}]4363\AA\ could not be accurately measured
for the emitting material lying further to the east (spectra S7, S8),
but given the observed line widths, the line ratios
for the extended narrow emission are most likely to be similar, reflecting
AGN photoionization. To the west
of the host galaxy, the spectra extracted across the emission line
arc, i.e. the region of the EELR with the strongest impact from
jet--cloud interactions
(S1, S2 and S3) display a mixture of emission line ratios. S2 and S3,
while only having lower limits for the
[O\textsc{iii}]5007/[O\textsc{iii}]4363 line ratio, are more
consistent with a shock plus precursor photoionization region
ionization mechanism than with shock ionization alone.  S1 lies
beyond the radio lobe where jet--cloud interactions are not yet likely
to have affected the gas strongly; comparison with S2 and S3 reflects the impact
that such interactions have on the observed line ratios.  Overall,
these results are fully consistent with those determined from the
spectroscopic studies of VM99 and Clark et al (1997).

We will now consider the central regions of the EELR in more detail.

\subsubsection{The velocity curve of the inner 2$^{\prime\prime}$}

\begin{figure*}
\vspace{2.3 in}
\begin{center}
\includegraphics{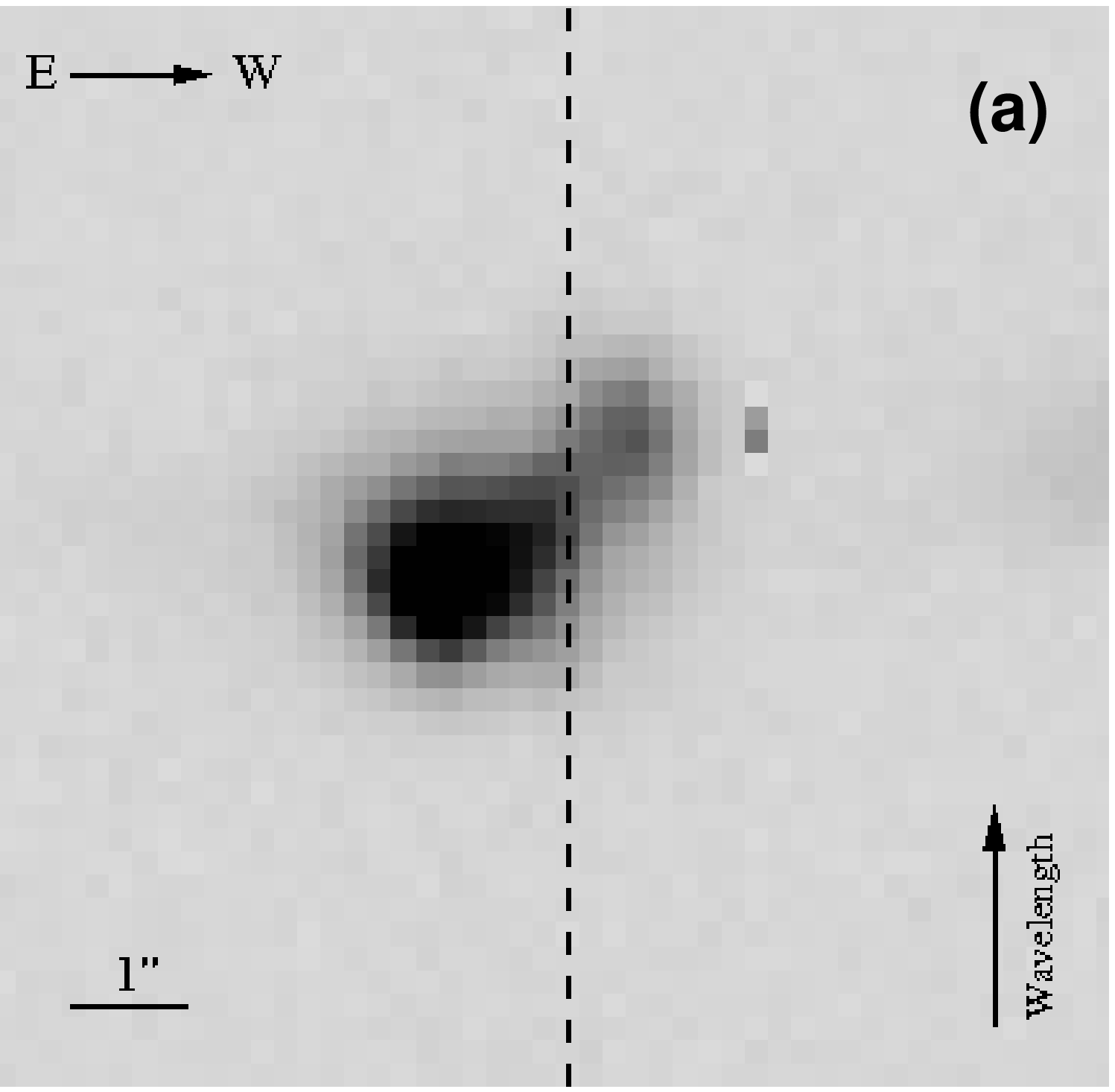} 
\includegraphics{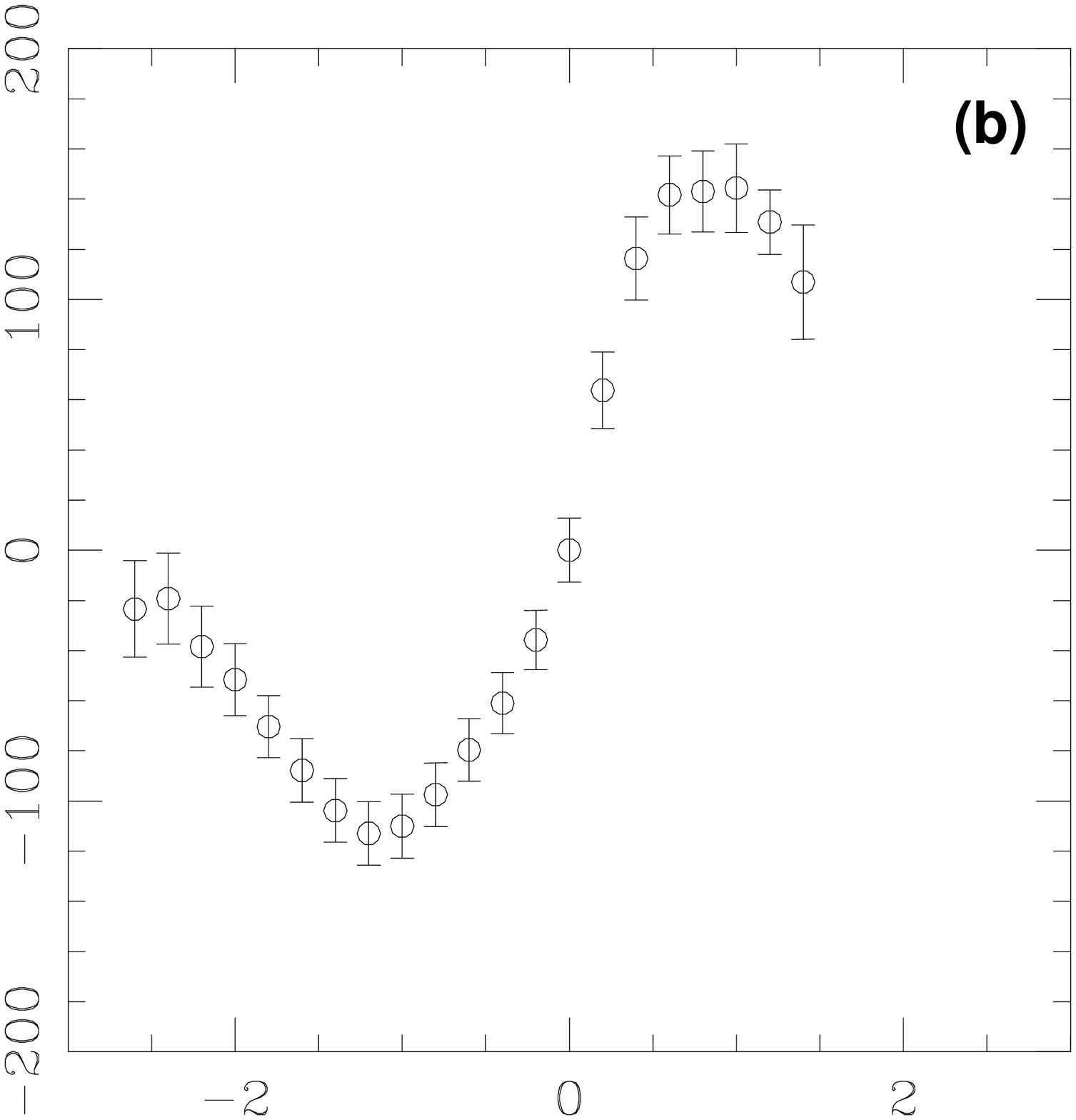}
\includegraphics{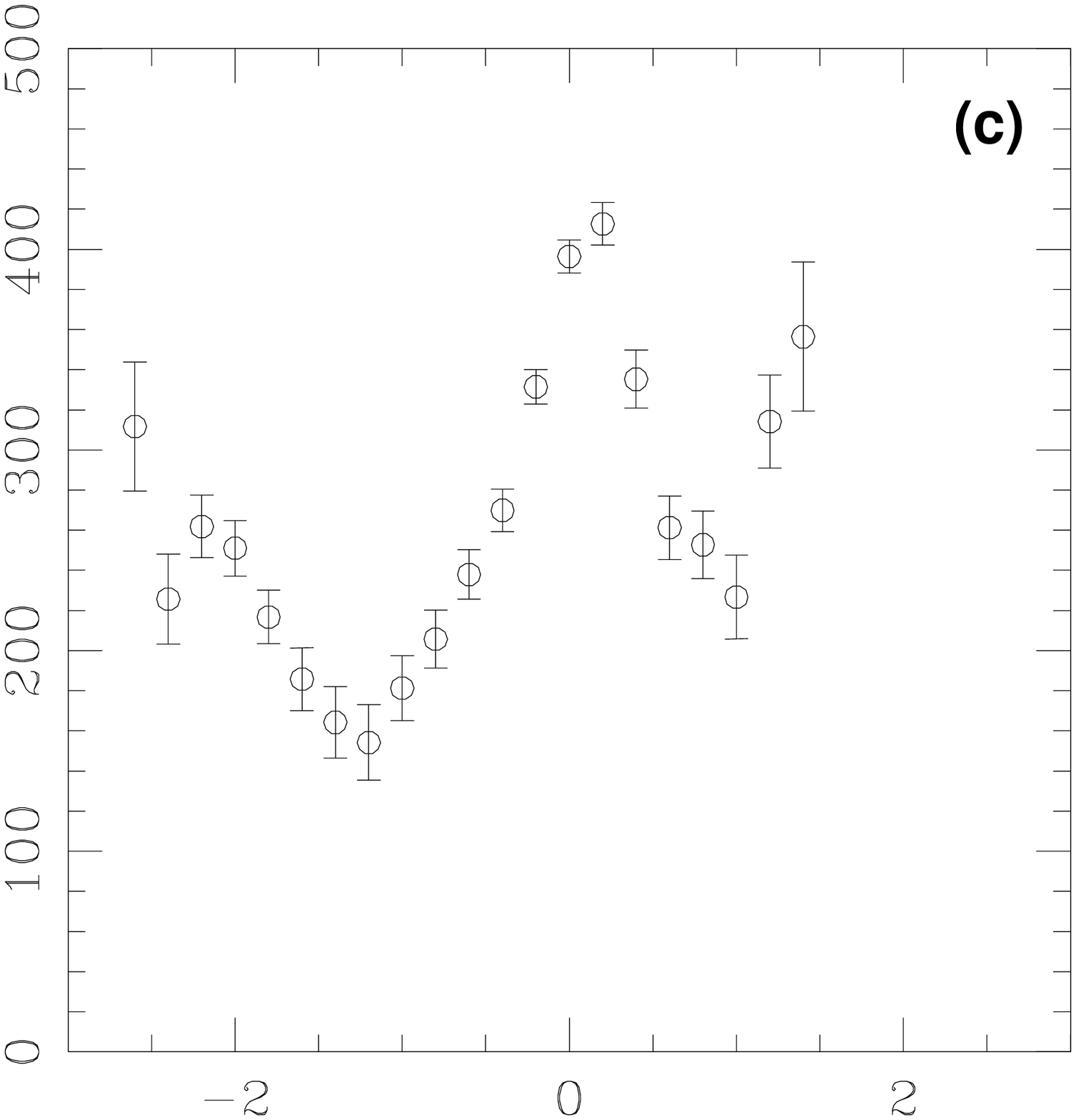} 
\end{center}
\caption{Left: Inner $4^{\prime\prime}$ ($\sim 18$kpc) region of the
  FORS1 spectrum (PA $69^{\circ}$),
  illustrating the continuum-subtracted [O\textsc{iii}]5007\AA\
  emission line (the continuum centroid location is indicated by the
  dashed line). Centre: [O\textsc{iii}]5007\AA\ emission line velocity
  shift ($\rm km\,s^{-1}$) as a function of offset (east to the left,
  west to the right, measured in arcsec) from the continuum
  centroid. Right:
  [O\textsc{iii}]5007\AA\ emission line FWHM ($\rm km\,s^{-1}$) as a function of offset (arcsec) from the continuum
  centroid. All 
  velocities are in the rest-frame of the galaxy.  As a reminder, 1
  arcsec is equivalent to 4.55kpc in our assumed cosmology.
\label{2250spec}}
\end{figure*}

The line-emitting gas close to the host galaxy has a number of
interesting features.  In Fig.~\ref{2250spec}, we display a
continuum-subtracted image of the [O\textsc{iii}]5007\AA\ emission
line from the two-dimensional FORS1 spectrum.  Also displayed are the
emission line velocity shifts and FWHM, evaluated on a pixel-by-pixel
basis, with the spatial zero taken at the position of the continuum
centroid. The FWHM (after correcting for instrumental broadening)
varies in the range $\sim 150-410 \rm km\,s^{-1}$, with a maximum at
the position of the continuum centroid (possibly due to unresolved
rotation in the central regions).  The line emission is
asymmetric at some spatial offsets, where the line profile is better
reproduced by multiple Gaussian components.  While the line broadening and
asymmetries might be expected to distort the
velocity profile, the fact that it remains remarkably smooth and does
not display sharp changes at the position of maximum line broadening
suggests that the overall kinematic pattern of the gas is not severely
distorted.

\begin{figure}
\vspace{2.15 in}
\begin{center}
\includegraphics{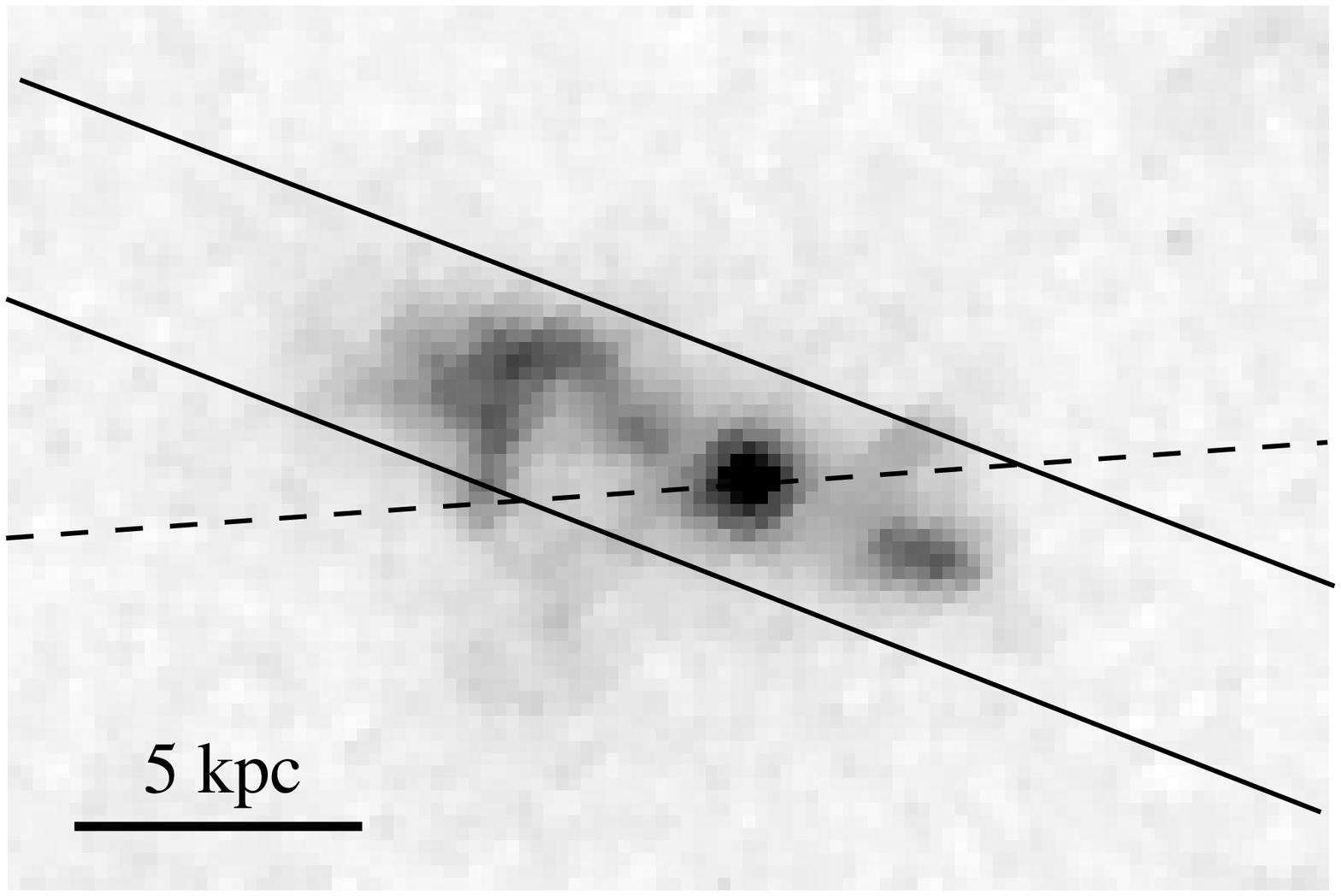} 
\end{center}
\caption{Drizzled WFPC2 [O\textsc{iii}]5007\AA\ emission-line image of the
  central regions of the EELR surrounding PKS2250-41, illustrating the
  strongest line emission.  The position of
  the FORS1 slit is marked by the angled lines. The radio source
  axis is aligned approximately east-west, and is marked by the dashed
  line.  In this figure, north is to the top and east to the left.
\label{oiii_closeup}}
\end{figure}

At this point, it is interesting to examine in detail the high surface
brightness emission line features present in the HST imaging
observations of T05.
Figure~\ref{oiii_closeup} displays a continuum-subtracted [O\textsc{iii}]5007\AA\ emission
line image of the EELR, with the contrast
levels adjusted to best display the morphology of the high surface
brightness features in the
central $3-4$arcsec.
Line emitting material extends out from the galaxy nucleus along
the same position angle as the spectroscopic slit, i.e. in the
direction of the companion galaxy; the eastern
emission is brighter and has a less regular shape.  Fainter line
emission surrounds these structures, filling the spectroscopic slit.

It is possible that the difference between the seeing ($\sim
0.56^{\prime\prime}$) and the slit width ($\sim 1.3^{\prime\prime}$) 
could have introduced an erroneous instrumental profile: in the case
of an unresolved emission feature which does not fill the slit
completely, the instrumental profile would be $2.54$\AA\ rather than
the $5.9 \pm 0.2$\AA\ measured from the sky and arc lines.  Although
the emission line imaging observations confirm that line emission does
indeed fill the FORS1 slit (Fig.~\ref{oiii_closeup}), the brightest features clearly occupy a
more confined region within the bounds of the slit, and are therefore
a potential problem.  While this could potentially lead to artificial velocity
shifts, the close similarity between the kinematics derived from the FORS1
observations and the VIMOS observations suggest that there is no
problem in this respect.

There are several possible interpretations for the observed gas
kinematics. On the basis of the FORS1 spectrum alone, the gas
kinematics are consistent with a $\sim 15$kpc diameter rotating disk,
and an implied lower limit to the dynamical mass of $M sin i^{-1}
\sim 2 \times
10^{10} \rm M_{\odot}$ within a 1 arcsec/4.55kpc radius.  Although
this is a lower limit, the morphology of the EELR
(Fig.~\ref{oiii_closeup}) would suggest a close to edge-on orientation
for any disk system present. Alternatively, the observed velocities
could be caused by other means, with the observed line
emission originating primarily from the edges of an ionization
cone. Close inspection of Fig.~\ref{oiii_closeup} suggests that there
is a dearth of emission {\it directly} along the east-west aligned
radio source axis, suggesting a hollowed-out region between the
brightest emission extending NE-SW and the fainter emission extending
SE-NW.  This is similar to the conic features observed in Cygnus A
(Taylor et al 2003; Tadhunter et al 1999; Jackson et al 1998).  We
note further that that the observed velocity amplitude displayed in
Fig.~\ref{2250spec} is much less than that observed in other galaxies
with gas undergoing regular rotation (e.g. Tadhunter et al 1989, Baum
et al 1990).  Other alternative explanations which should be considered
are inflows or outflows imperfectly aligned with the radio source axis
(see Fig.~\ref{oiii_closeup}).

\subsection{IFU Spectroscopy}

As the EELR of PKS2250-41 are known to be so complex, integral field
spectroscopy is a particularly effective means of disentangling the
varied kinematic components.  In this section, we will discuss the
different properties of the EELR in turn, focusing on the EELR
morphology, kinematics and ionization state.  In order to best
illustrate the different features of the EELR, we present our IFU data
in several different ways.

\begin{figure}
\vspace{4.65 in}
\begin{center}
\includegraphics{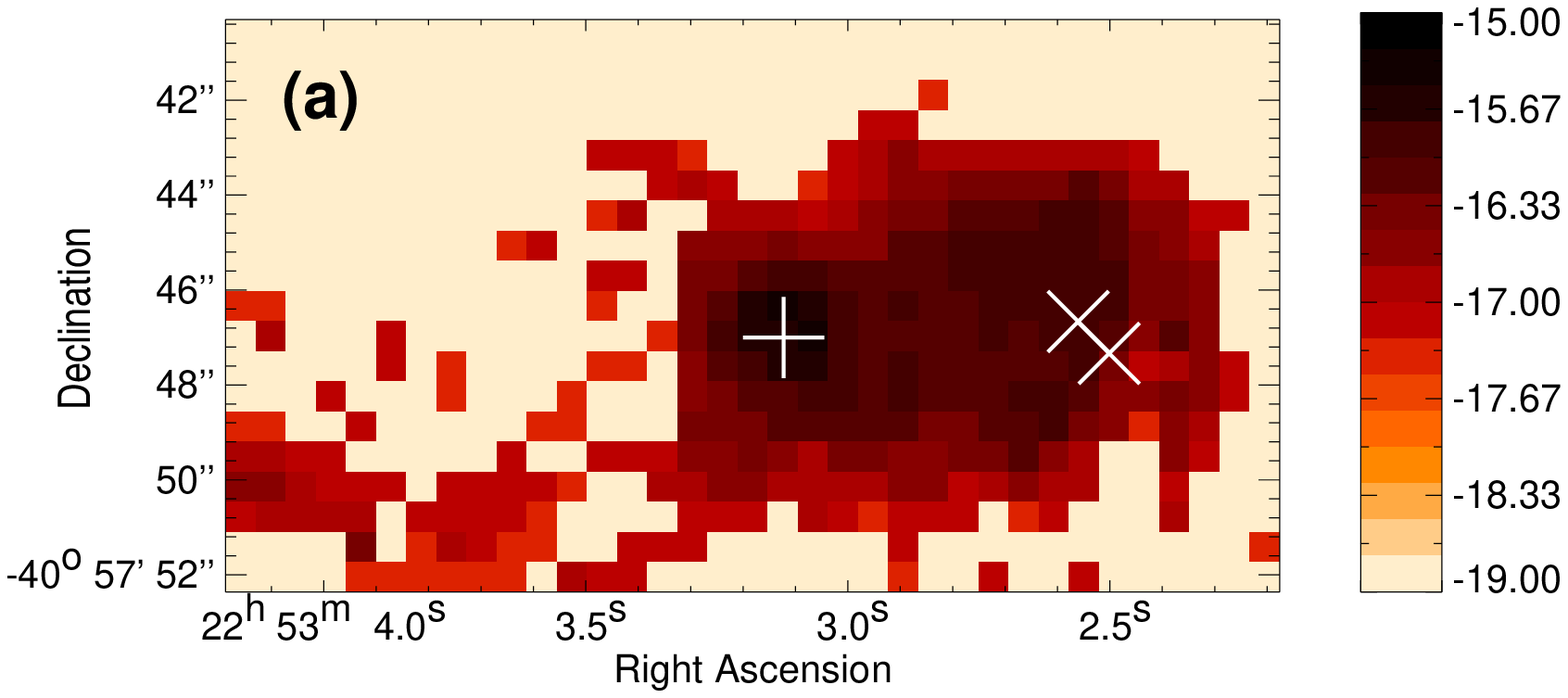}
\includegraphics{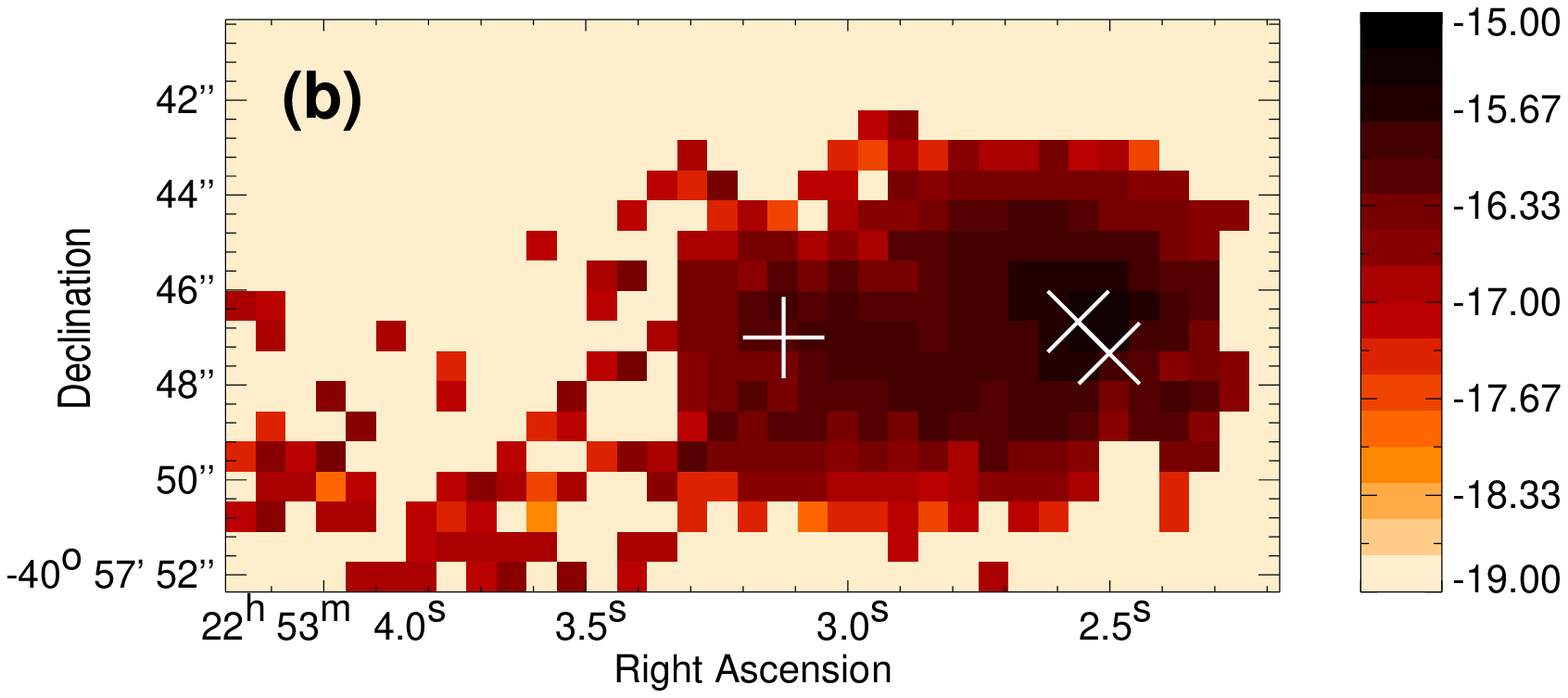}
\includegraphics{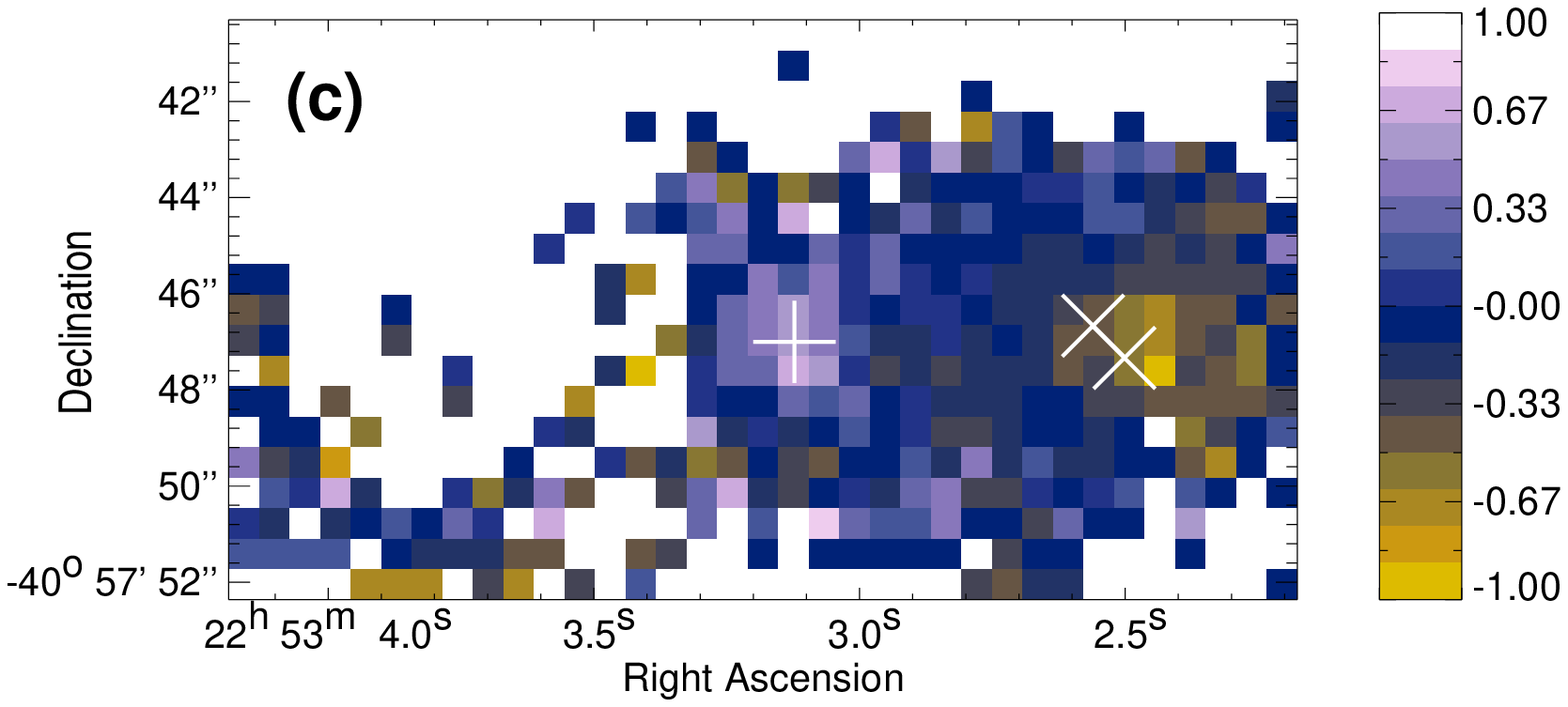}
\end{center}
\caption{From top: (a) extracted [O\textsc{iii}]5007\AA\ line flux,
  (b) extracted [O\textsc{ii}]37277\AA\ line flux, and (c)
  [O\textsc{iii}]5007\AA/[O\textsc{ii}]37277\AA\ emission line
  ratio derived from the VIMOS data cube. These results are based on single gaussian fits.  The position of the host galaxy
  centroid is marked in all 
  frames by `+', and the position of the two western-lobe radio
  hotspots by `x'. For comparison with the
  radio and continuum structures, we refer the reader back to
  Fig.~\ref{Fig: radio}b. The grey-scale bars for
  the flux images are
  logarithmic, in units of $\rm erg s^{-1} cm^{-2}$.  The scale bar
  for frame (c) is logarithmic and dimensionless.  North is to the top
  and east to the left, the pixel scale is $0.67^{\prime\prime}$/pixel
  (giving a total field of view of $\sim23.5 \times 13$ square arcsec,
  or $\sim 110 \times 60 \rm kpc^2$ at a redshift of $z=0.308$).
\label{IFU_fluxes}}
\end{figure}

The figures in this section (\ref{IFU_fluxes}, \ref{IFU_kinematics} and \ref{IFU_regions}) concentrate on our chosen
region of the IFU data cube (illustrated in Fig.~\ref{Fig: radio}b),
which covers the majority of the radio cocoon. 
In order to study the EELR properties in greater detail, we have developed IDL routines
to model the line emission on a fibre-by-fibre basis.  For each fibre
spectrum, we locate a selected emission line and fit it with one or more Gaussian
components plus a continuum, extracting the resulting modelled emission-line flux, FWHM and velocity offset
relative to the systemic redshift of PKS2250-41.
Fig.~\ref{IFU_fluxes} displays the variation in the fluxes of the
[O\textsc{iii}]5007\AA\ and [O\textsc{ii}]3727\AA\ emission lines
across the EELR, and also the changing
[O\textsc{iii}]5007\AA/[O\textsc{ii}]3727\AA\ line ratio for the same
region.   Fig.~\ref{IFU_kinematics} displayes the kinematic properties
of the modelled [O\textsc{iii}]5007\AA\ and [O\textsc{ii}]3727\AA\ emission lines.

\subsubsection{The EELR morphology}

The EELR surrounding PKS2250-41 displays a number of different
features.   
The brightest [O\textsc{iii}] line emission is located close to the host galaxy and is approximately
co-spatial with the continuum centroid.  This is in good agreement with previous
spectroscopic studies of this source (VM99, Clark et al 1997). However, the line emission is noticeably 
elongated, with the long axis orientated along approximately the same
angle as the FORS1 slit and the filamentary line emission displayed in
Fig.~\ref{oiii_closeup}.  Although faint line emission is observed
along this PA for a considerable distance in our FORS1 spectrum, the
IFU data only detect this emission relatively close to the host
galaxy.  Faint emission is observed in the IFU data cube to the south
east of the host galaxy, coincident with some of the features observed
in the ground-based emission line imaging observations of Clark et al
(1997).  The powerful emission-line arc is also clearly detected in
our IFU data.  However, we also observe fainter emission filling the
region between the host galaxy and the emission-line arc.

\subsubsection{Ionization of the emission line gas.}

As Fig.~\ref{IFU_fluxes}c demonstrates, the ionization state of the
gas varies quite strongly across the EELR.  The ratio of
[O\textsc{iii}]5007\AA/[O\textsc{ii}]3727\AA\ is at its highest close
to the host galaxy nucleus, consistent with AGN photoionization
dominating in this region.  The [O\textsc{iii}]5007\AA/[O\textsc{ii}]3727\AA\ 
line ratio is much lower towards the western emission line arc, reaching a minimum
level at the position of the radio source hotspot, where jet-induced
shocks are likely to dominate. Westwards of the most intense emission
in the arc the oxygen line ratio rises once again. It is also noteworthy that a lower
ionization state is observed {\it along} the radio source axis, with (relatively)
stronger [O\textsc{iii}] emission to either side.  Again, these
results are consistent with those of VM99 and Clark et al (1997).

\subsubsection{EELR kinematics}

\begin{figure*}
\vspace{3. in}
\begin{center}
\includegraphics{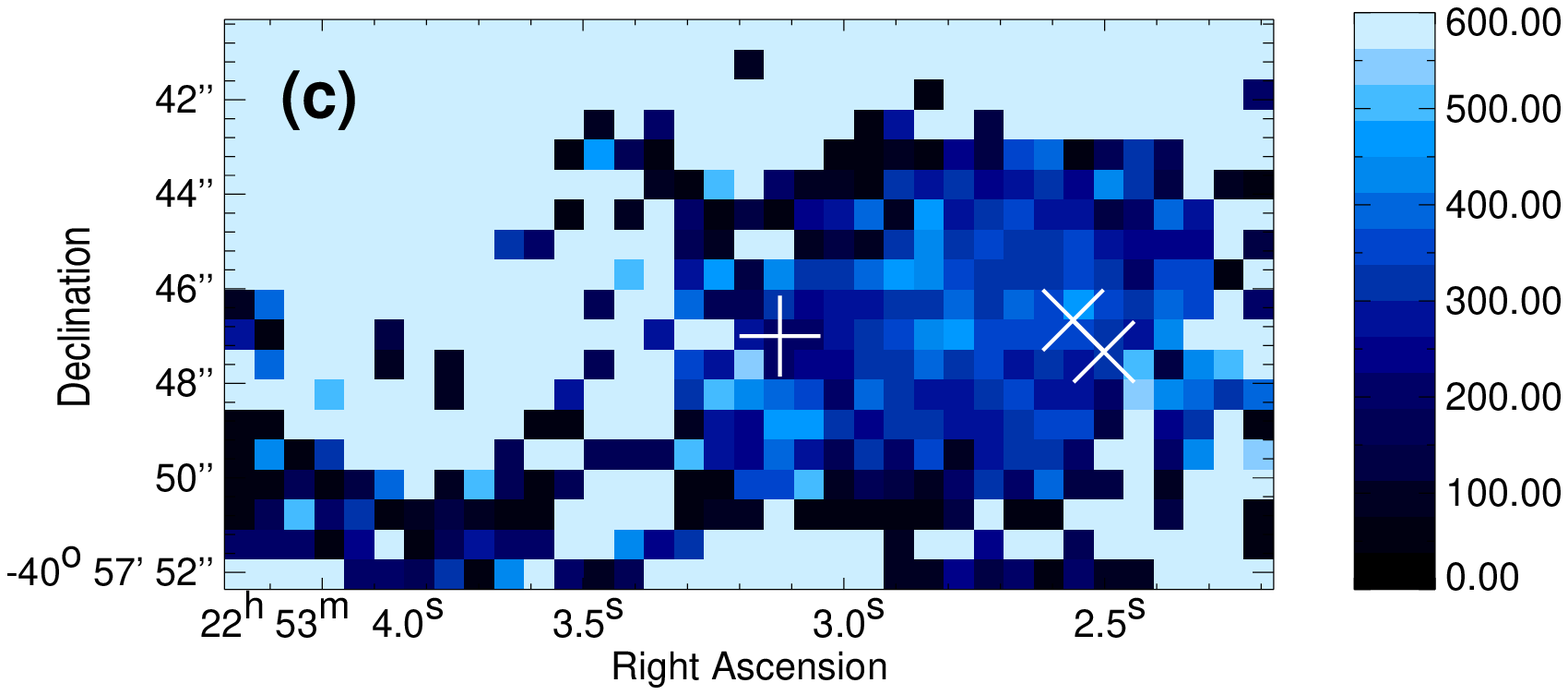}
\includegraphics{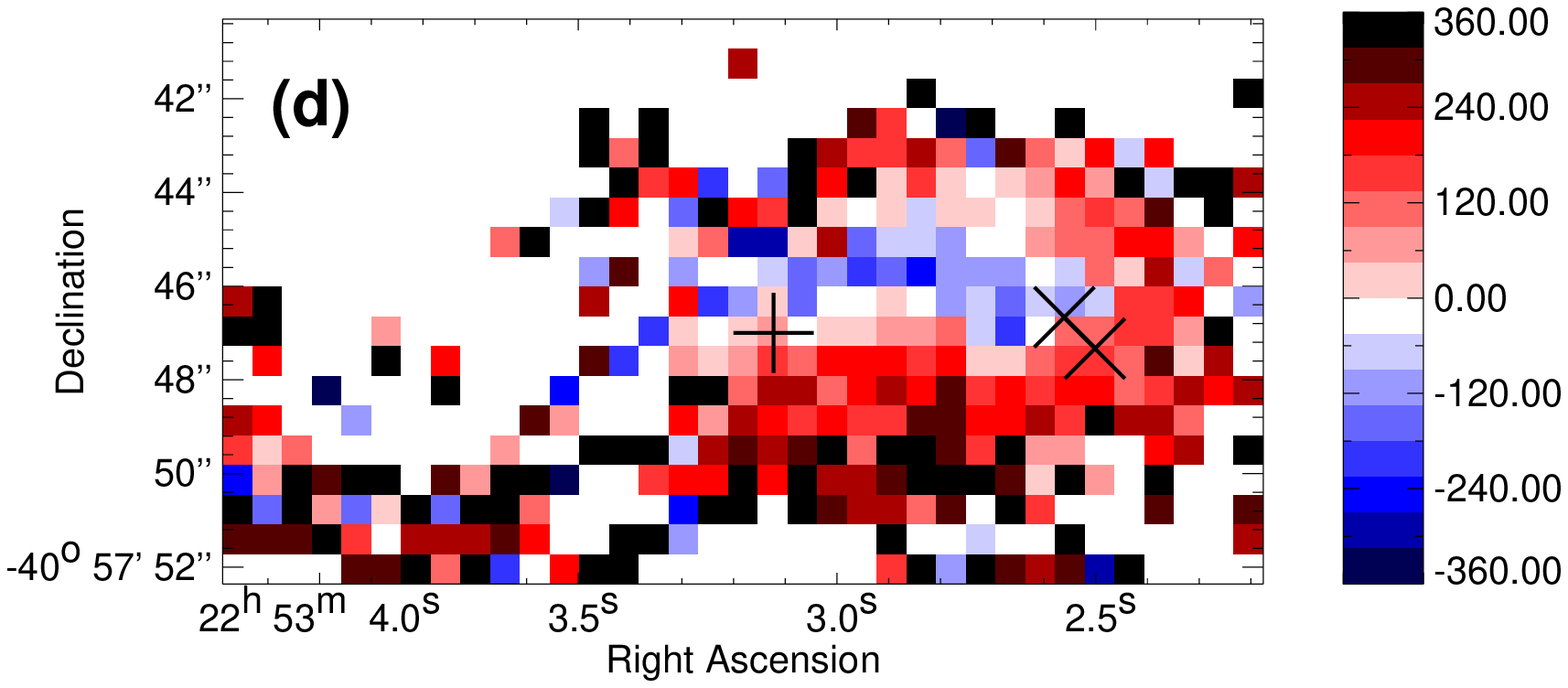}
\includegraphics{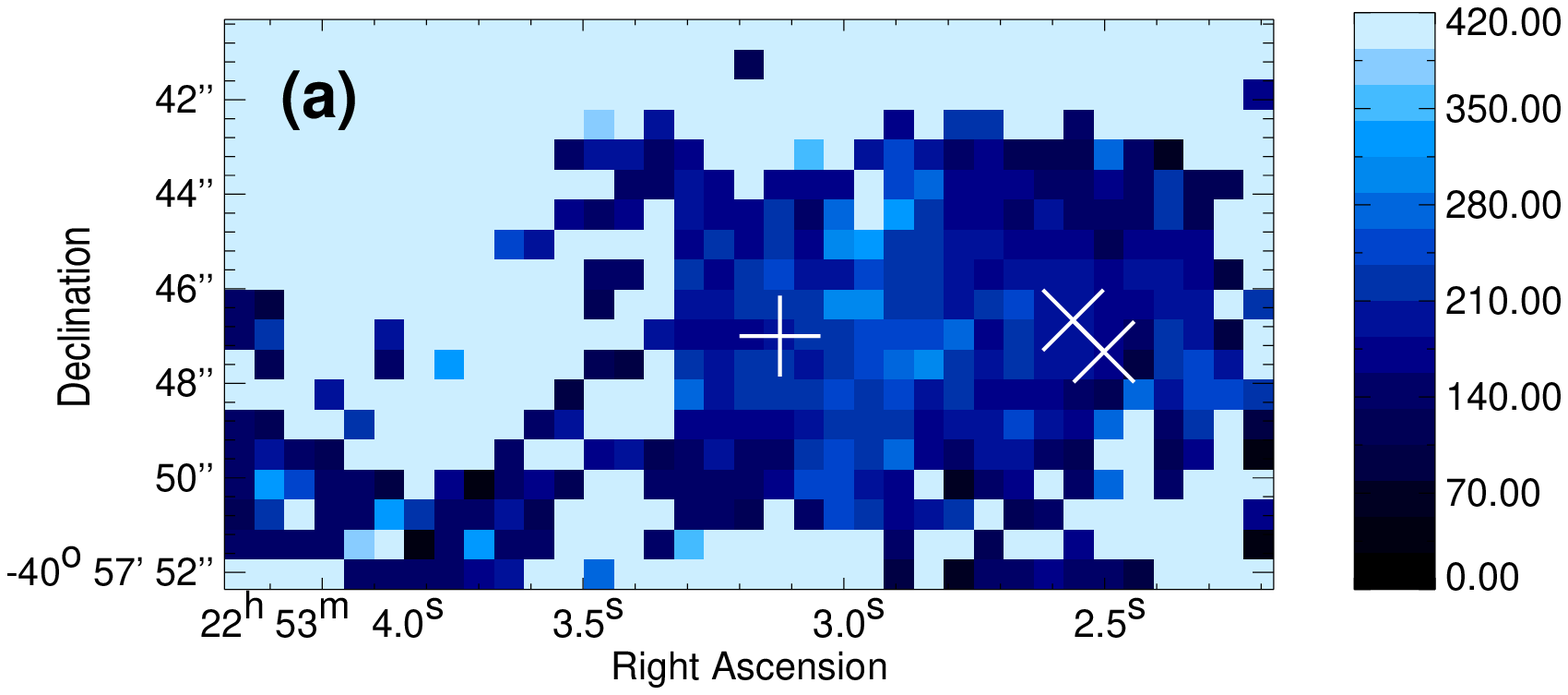}
\includegraphics{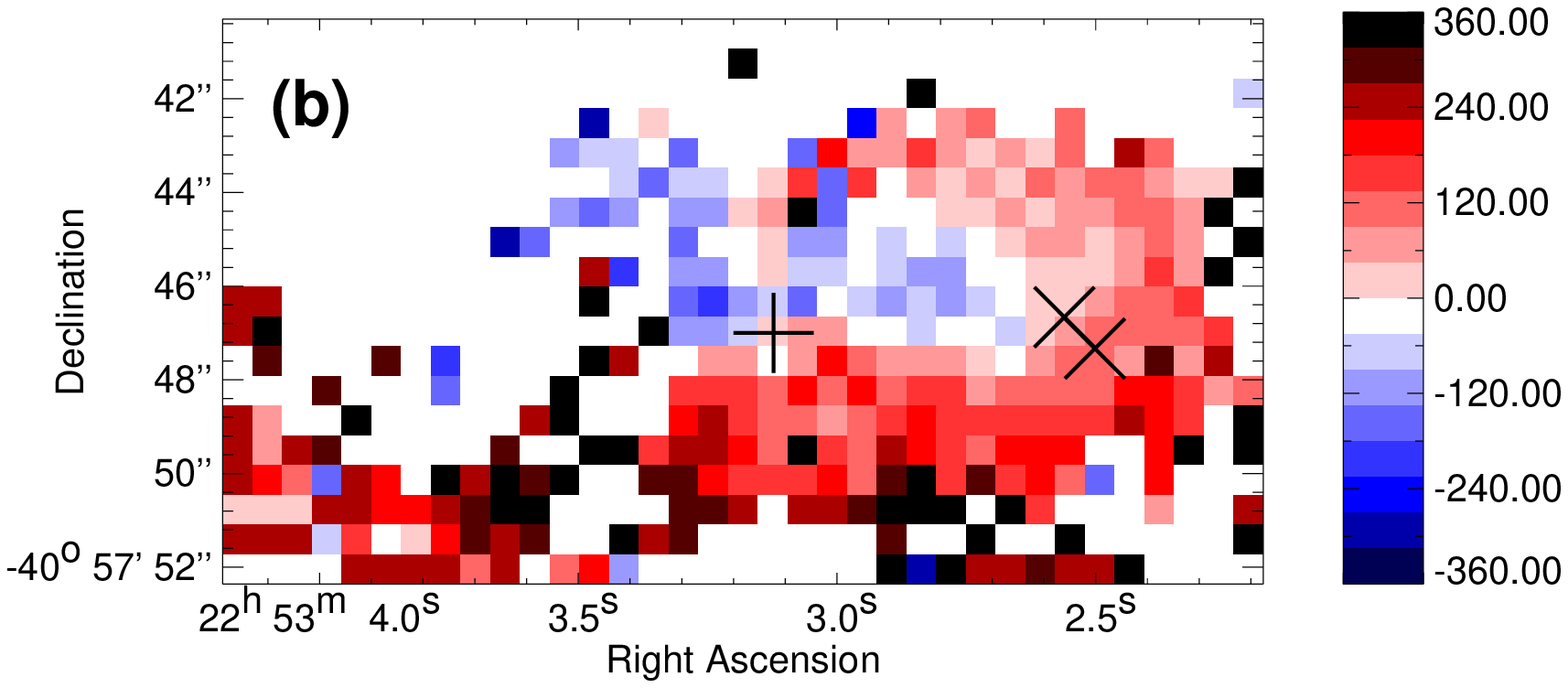}
\end{center}
\caption{IFU emission line kinematics. (a - top left)
  [O\textsc{iii}]5007\AA\ FWHM line width (corrected for instrumental
  broadening). (b - lower left) velocity shift of
  [O\textsc{iii}]5007\AA\ relative to $z=0.3074$. (c - top right) 
  [O\textsc{ii}]3727\AA\ FWHM line width (corrected for instrumental
  broadening). (d - lower right) velocity shift of
  [O\textsc{ii}]3727\AA\ relative to $z=0.3074$. These results are
  based on single gaussian fits.  The position of the host galaxy
  centroid is marked in all 
  frames by `+', and the position of the two western-lobe radio
  hotspots by `x'. In these plots, we have
  masked the fibres with an [O\textsc{iii}]5007\AA] emission line
  flux below $1 \times 10^{-18} \rm erg s^{-1} cm^{-2}$.  The colour
  bars denote velocities in units of $\rm km s^{-1}$.  North is to the top
  and east to the left, the pixel scale is $0.67^{\prime\prime}$/pixel
  (giving a total field of view of $\sim23.5 \times 13$ square arcsec,
  or $\sim 110 \times 60 \rm kpc^2$ at a redshift of $z=0.3074$.
\label{IFU_kinematics}}
\end{figure*}

One feature which is immediately obvious from our IFU data is the
velocity shift across the western emission line arc; as
Fig.~\ref{IFU_kinematics} illustrates, the material to the north of
the arc is clearly blueshifted with respect to the emitting material
in the southern portions of the arc, and the pixel to pixel variation
is generally quite smooth.  This is consistent with the data obtained
from the current and previous long-slit spectroscopic studies, but the observed
velocity shift is far more readily apparent in the IFU data.  The north-south velocity shift across the EELR
is in the range 100-300$\rm km\,s^{-1}$. Such a
general north-south divide between redshifted and blueshifted material
is suggestive of bulk rotation, though as noted in section 3.1.1, the
observed velocities are somewhat on the low side.   

In terms of the widths and profiles of the emission lines, the
broadest and most complex emission is observed along the radio source
axis, between the host galaxy centroid and the radio source hotspot.
The emitting material towards the outer edge of the western arc
displays relatively narrow emission lines, as does the emission
extending into the eastern radio lobe.

\begin{figure}
\vspace{1.8 in}
\begin{center}
\includegraphics{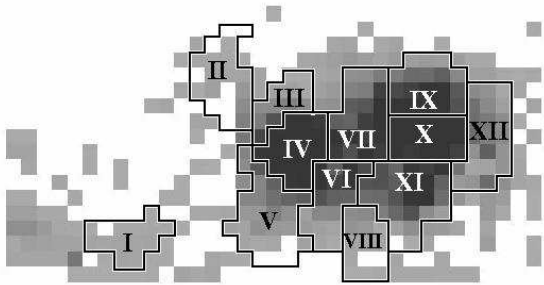}
\end{center}
\caption{[O\textsc{iii}]5007\AA\ IFU flux image illustrating the twelve
  multi-fibre regions extracted as part of our higher signal-to-noise
  analysis of the EELR. Regions have been selected according to
  similar kinematic properties of adjacent fibres.   North is to the top
  and east to the left, the pixel scale is $0.67^{\prime\prime}$/pixel
  (giving a total field of view of $\sim23.5 \times 13$ square arcsec,
  or $\sim 110 \times 60 \rm kpc^2$ at a redshift of $z=0.308$. The
  galaxy nucleus lies within region IV, while the bright emission line
  arc to the west of the galaxy is covered by regions IX, X and XI.
\label{IFU_regions}}
\end{figure}

In order to analyse our IFU spectra at a higher signal--to--noise
level and address the issue of superposed velocity structures,
we have extracted and combined the spectra of fibres covering specific
emission regions, and those which display similar kinematic
properties.  The resulting regions, labelled numerically from I to
XII, are displayed in
Fig.~\ref{IFU_regions}. 
We have re-analysed the
[O\textsc{iii}]5007\AA\ line widths and velocity shifts, fitting up to
three Gaussians to the combined emission line data.  The results of
our kinematic analysis are consistent between different lines, and with
the single-fibre data displayed in Fig.~\ref{IFU_kinematics}; we 
tabulate the results for [OIII]4959,5007 in Table 2.  These results
are also in good agreement with those of  the
current FORS1 spectroscopy and the 
earlier spectroscopy of Clark et al (1997), Tadhunter et al (1994) and VM99.

\begin{table}
\caption{Kinematics of the multi-fibre regions illustrated in
  Fig.~\ref{IFU_regions}. Velocity shifts are relative to
  the typical value of $z=0.3074$.}
\begin{tabular}{lccc}
{Region} & {FWHM}  & {Velocity Shift} & {Reduced $\chi^2$}\\
& ($km\,s^{-1}$) & ($km\,s^{-1}$) & \\\hline
I - single& $ 197 \pm 32$ & $ 278 \pm 35 $  & 1.54\\
II - single& $ 144 \pm 32$ & $-136 \pm 35 $  & 1.60\\
III - single& $ 436 \pm 14$ & $ -69 \pm 35 $  & 2.52\\
IV - single& $ 438 \pm  2$ & $ -10 \pm 34 $  & 5.93\\
IV - double& $ 381 \pm  3$ & $  -8 \pm 34 $  & 5.52\\
               & $ 700 \pm 16$ & $ -21 \pm 35 $  &\\
V - single& $ 335 \pm  8$ & $ 162 \pm 34 $  & 3.59\\
V - double& $ 138 \pm 14$ & $ 173 \pm 34 $  & 2.26 \\
               & $ 819 \pm 44$ & $  92 \pm 40 $  &\\
VI - single & $ 479 \pm 11$ & $ 130 \pm 34 $  & 1.60\\
VI - double & $ 409 \pm 12$ & $ 155 \pm 34 $  & 1.55\\
       & $ 668 \pm 55$ & $  21 \pm 42 $  &\\
VII - single & $ 442 \pm  6$ & $ -55 \pm 34 $  & 2.58\\
VII - double & $ 169 \pm 13$ & $ -37 \pm 34 $  & 1.60\\
       & $ 620 \pm 12$ & $ -74 \pm 34 $  &\\
VIII - single & $ 410 \pm 59$ & $ 295 \pm 41 $  & 2.33\\
IX - single & $ 204 \pm  5$ & $  57 \pm 34 $  & 2.17\\
IX - double & $ 172 \pm  5$ & $  59 \pm 34 $  & 1.95\\
               & $ 819 \pm 130$& $ -85 \pm 69 $  &\\
X - single & $ 319 \pm  4$ & $  22 \pm 34 $  & 5.86\\
X - double & $ 174 \pm  6$ & $  33 \pm 34 $  & 1.07\\
       & $ 940 \pm 32$ & $ -93 \pm 37 $  &\\
X - triple & $ 122 \pm  9$ & $  33 \pm 34 $  & 0.97\\
       & $1037 \pm 184$& $-619 \pm 94 $  &\\
       & $ 699 \pm 11$ & $  -4 \pm 34 $  &\\
XI - single & $ 327 \pm  6$ & $ 129 \pm 34 $  & 2.66\\
XI - double & $ 215 \pm  7$ & $ 129 \pm 34 $  & 1.72\\
       & $ 819 \pm 45$ & $ 120 \pm 40 $  &\\
XII - single & $ 274 \pm 12$ & $ 122 \pm 34 $  & 2.17\\

\end{tabular}
\end{table}

The narrowest emission line components are observed in the eastern radio lobe;
the line emission from regions I, II and V has instrumentally
corrected component line widths of the
order of 130-200$\rm km\,s^{-1}$.  However, even at these low FWHM
values the velocity shift between the northern and southern parts of
the EELR is clear. Similarly quiescent material with relatively narrow instrumentally
corrected line
widths (100-200$\rm km\,s^{-1}$) is also observed in the western radio
lobe, between the host galaxy and the emission line arc (regions VII,
IX, X, XI).  

The next notable feature is line emission with FWHM roughly of the order of
$400 \rm km\,s^{-1}$, which is observed in several consecutive
regions: III, IV, VI, VIII.  Interestingly, the emission from these regions lies along
the same PA as the bright emission line filament illustrated in
Fig.~\ref{oiii_closeup} and the FORS1 slit; our measured FWHM and
velocity shifts for
these regions are also consistent with those displayed in
Fig.~\ref{2250spec}.  However, as the spectra for regions III and VIII
have lower signal-to-noise levels than some of the other regions, we
cannot conclusively rule out a combination of broader and narrower
components; such degeneracies are a common issue in the analysis of
emission line profiles.  This may also be the case for region XII,
where the measured FWHM of 274$\rm km\,s^{-1}$ could perhaps be better
explained by broader emission associated with the arc lying adjacent
to more quiescent material. The emission from region XII lies beyond
the main emission line
arc and the western radio lobe and is at lower signal to noise than
the inner regions of the EELR.  Although a weak velocity gradient is
observed across this region (Fig.~\ref{IFU_kinematics}b), we cannot
conclusively confirm that the same apparent pattern of rotation holds
true in this region as well as in the inner EELR.

\begin{figure}
\vspace{8.02 in}
\begin{center}
\includegraphics{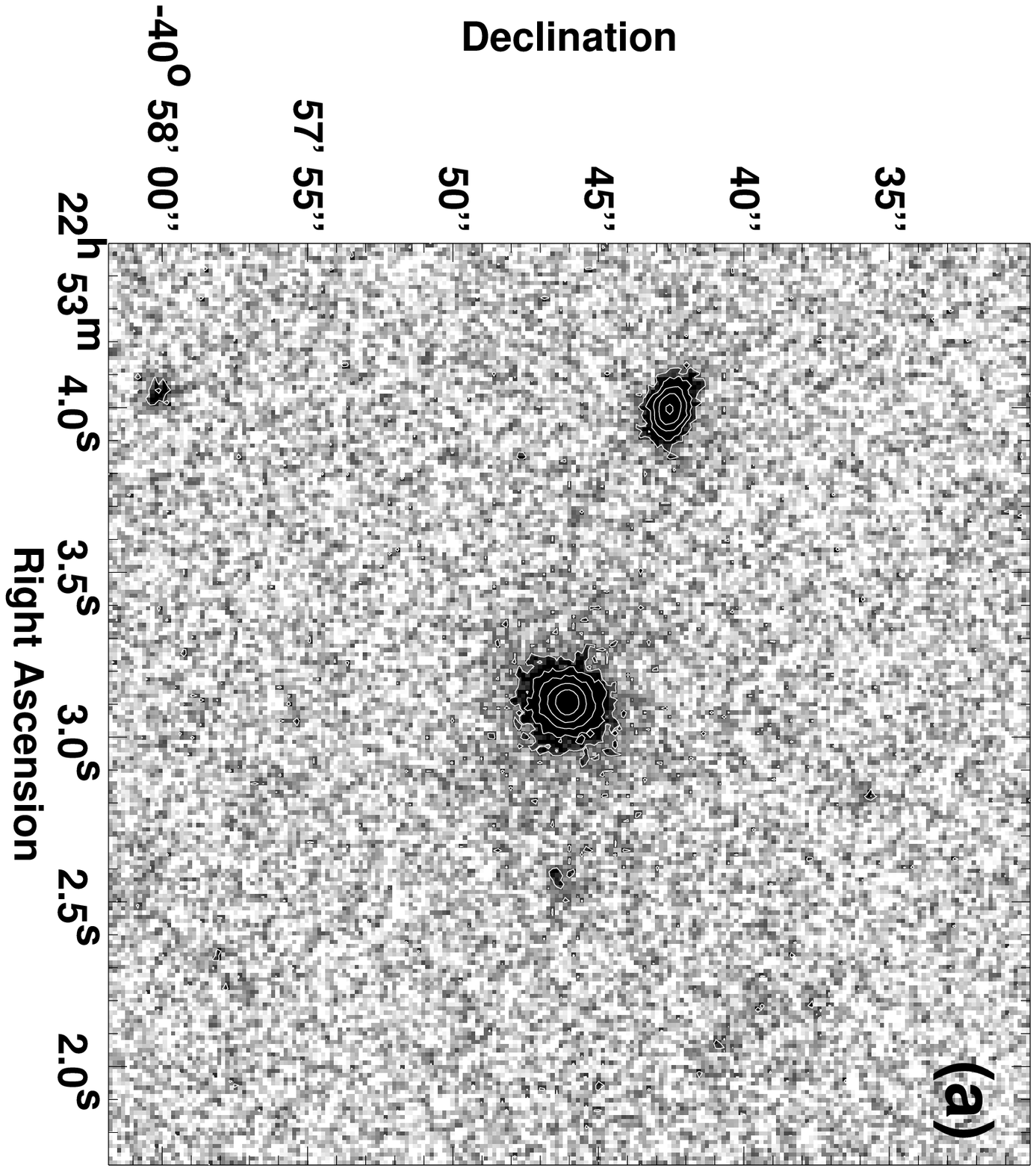}
\includegraphics{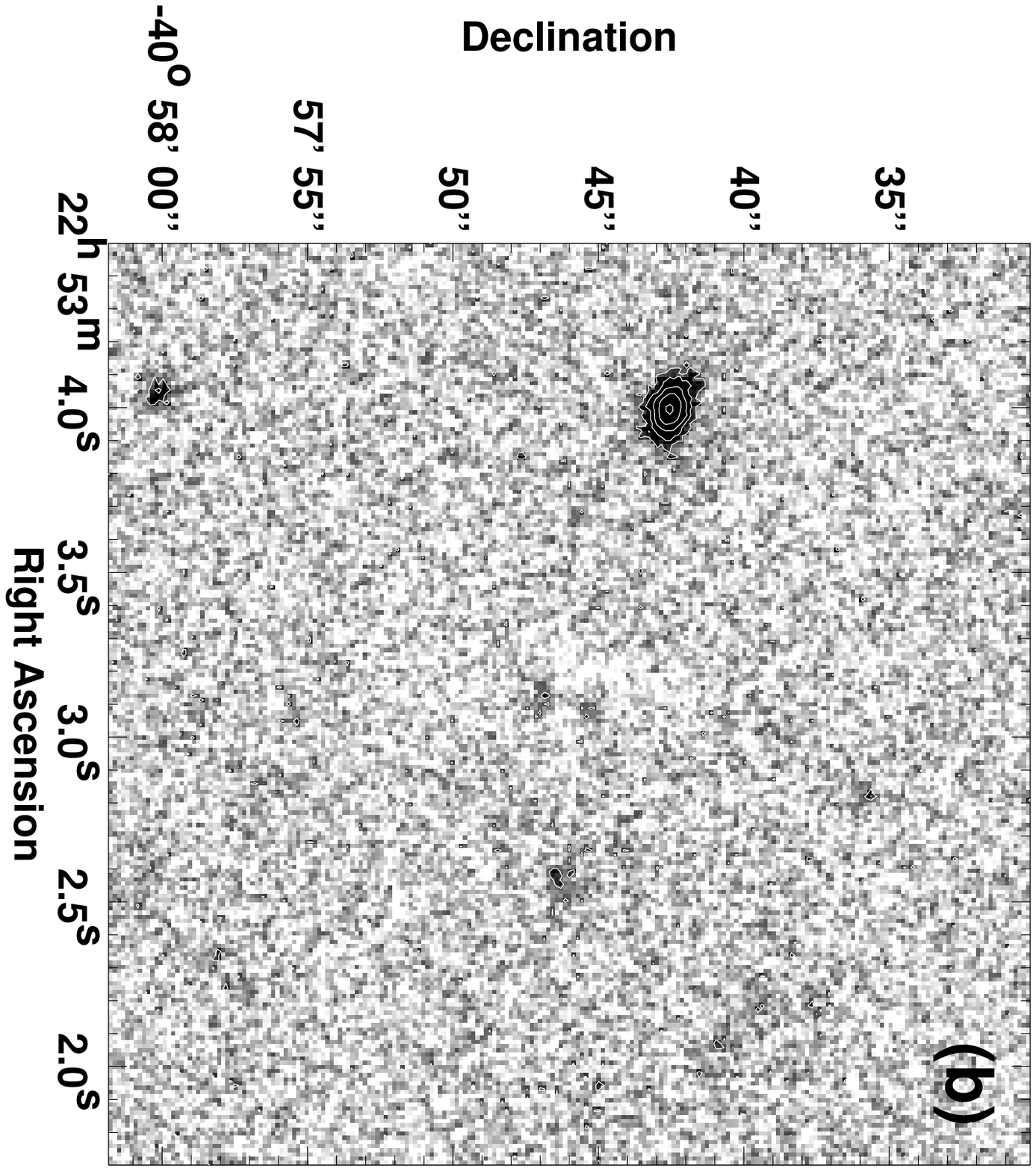}
\includegraphics{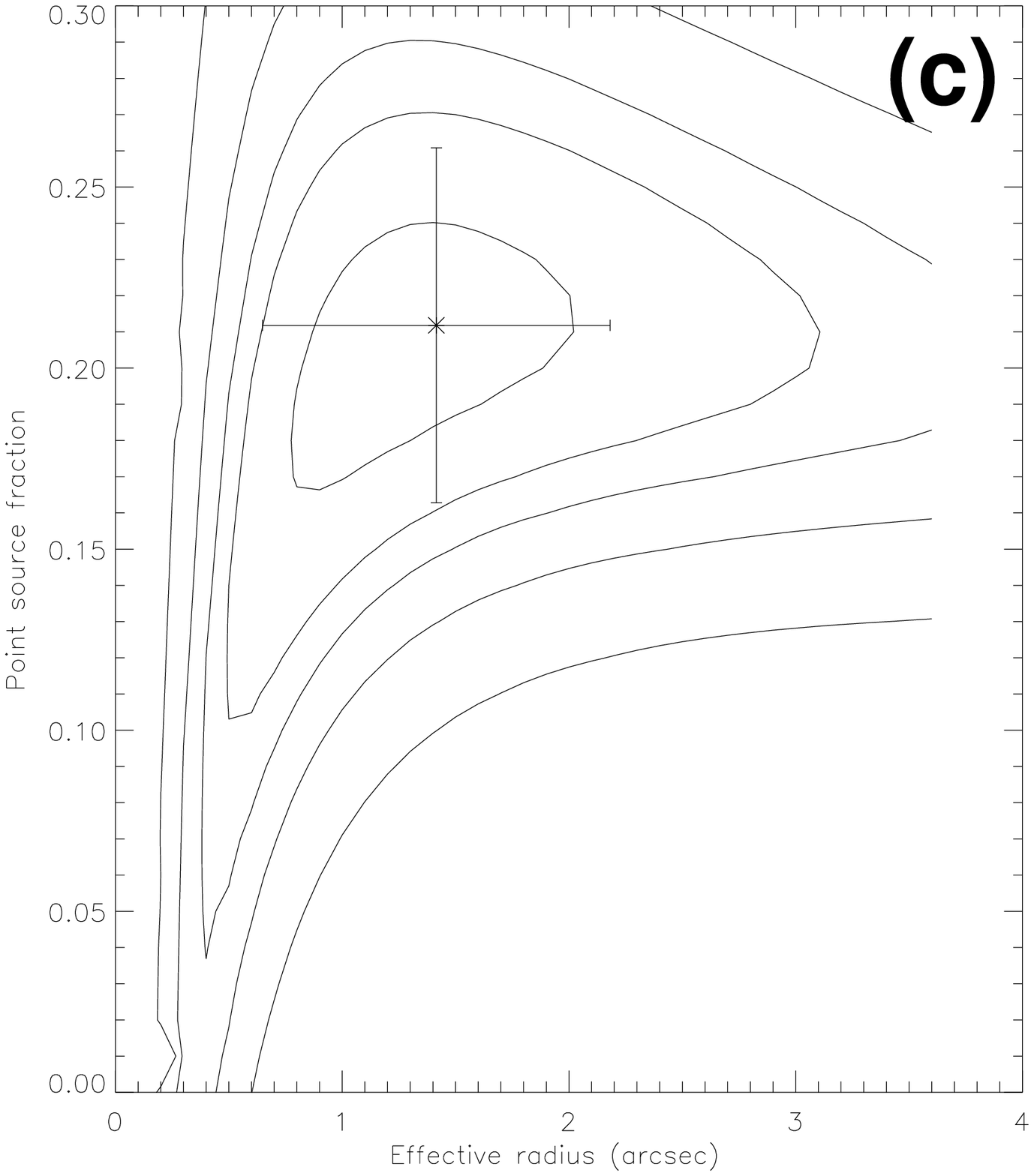}
\end{center}
\caption{Host galaxy morphology fits for PKS2250-41.  The $K-$band
   galaxy image is presented in frame (a), with the residuals after
   best-fit model galaxy subtraction displayed 
   in frame (b). Frame (c) illustrates the  1, 2, 3, 5 and 10-$\sigma$
   contours for the reduced $\chi^2$, plus the best-fit values and
   associated error bars for the
   point source contribution and effective radius.
\label{Fig: morph}}
\end{figure}

There is also some level of consistency between the regions lying
directly along the radio source axis (IV, VI, VII, IX, X, XI), all
of which feature a kinematic component with FWHM between $\sim
600-820\rm km\,s^{-1}$.  But once again, even these higher-velocity
features are subject to the same north-south gradient in relative
velocity.  This suggests that all of the different velocity components
present in the line-emitting gas are subject to the same underlying pattern of
rotation. 

Finally, the highest gas velocities are observed (as expected) in
close proximity to the radio source hotspot in the western radio lobe,
where we detect components with line widths of over 1000$\rm km\,s^{-1}$.

\subsection{The host galaxy}
Figure \ref{Fig: radio}b displays our recent $Ks-$band imaging
observations of PKS2250-41. The radio galaxy itself is an elliptical
with a total $K_S-$band magnitude (galactic 
extinction corrected and evaluated within a 10 arcsecond/45.5 kpc diameter
aperture) of $15.56 \pm 0.05$, placing it towards the fainter range of
host galaxy magnitudes in the radio source K-z relation (e.g. Inskip
et al 2002a; Inskip et al 2008 {\it in prep}). 

We have modelled the host galaxy morphology within a $32 \times 32\, \rm
arcsec^2$ region using the methods detailed in Inskip
et al. (2005), which can be briefly summarised as
follows.  The point spread function (PSF) for the SOFI field can be
accurately determined 
from unsaturated stars towards  the centre of the mosaic field; their 2-d
profile is extracted, normalised to unit flux, and used to generate an
average PSF profile. 
Once a good PSF had been obtained, de Vaucouleurs profile
 galaxy models were convolved with the PSF  and fitted to the surface profile of the galaxy,
using available least squares minimisation IDL
routines\footnote{\textsc{mp2dfunfit.pro}, part of Craig Markwardt's
\textsc{mpfit} non-linear least squares curve fitting package
available via http://astrog.physics.wisc.edu/$\sim$craigm/idl/fitting.html.}.  The free parameters for this
modelling are the galaxy flux,
centroid, effective radius, and fractional nuclear
point source contribution. Galaxy ellipticity was also allowed to
vary, but in the case of this object did not lead to significant
variation in the preferred effective radius, nuclear point source
contribution, or distribution of residual flux.

The modelled data and model-subtracted residuals are displayed in
Fig.~\ref{Fig: morph}, together with a plot of the variation in
reduced $\chi^2$ over the parameter space considered.  The resulting
best-fit parameters give an effective radius of $r_{\rm eff} = 1.42 \pm 0.77$
arcsec (equivalent to $6.47 \pm 3.51$kpc in our assumed cosmological
model), and an unresolved nuclear point source contribution of $21.2 
\pm 4.9$\%; the residuals are comparable to the background
noise. After subtraction of the point source component, the remaining
near-IR flux is comparable to that of a $\sim 1.4 \rm L_{\star}$ galaxy at
a redshift of $\sim 0.31$ (Willott et al 2003).

Assuming that the point source component is non-stellar in nature,
and that the bulk old stellar population of the galaxy formed at high
redshift (i.e. $z = 5-20$), the galaxy flux can be modelled by a
9.5Gyr old stellar population. We have convolved a 9.5Gyr GISSEL
spectral synthesis (Bruzual \& Charlot 2003) template spectrum with
the $K_S$-band filter profile.  Comparing the resulting flux with that
measured for PKS2250-41 (after subtraction of the point source
contribution measured via our modelling) suggests a total stellar mass
for the galaxy
of $2.1 \pm 0.5 \times 10^{11} \rm M_{\odot}$.

\begin{figure}
\vspace{6.2 in}
\begin{center}
\includegraphics{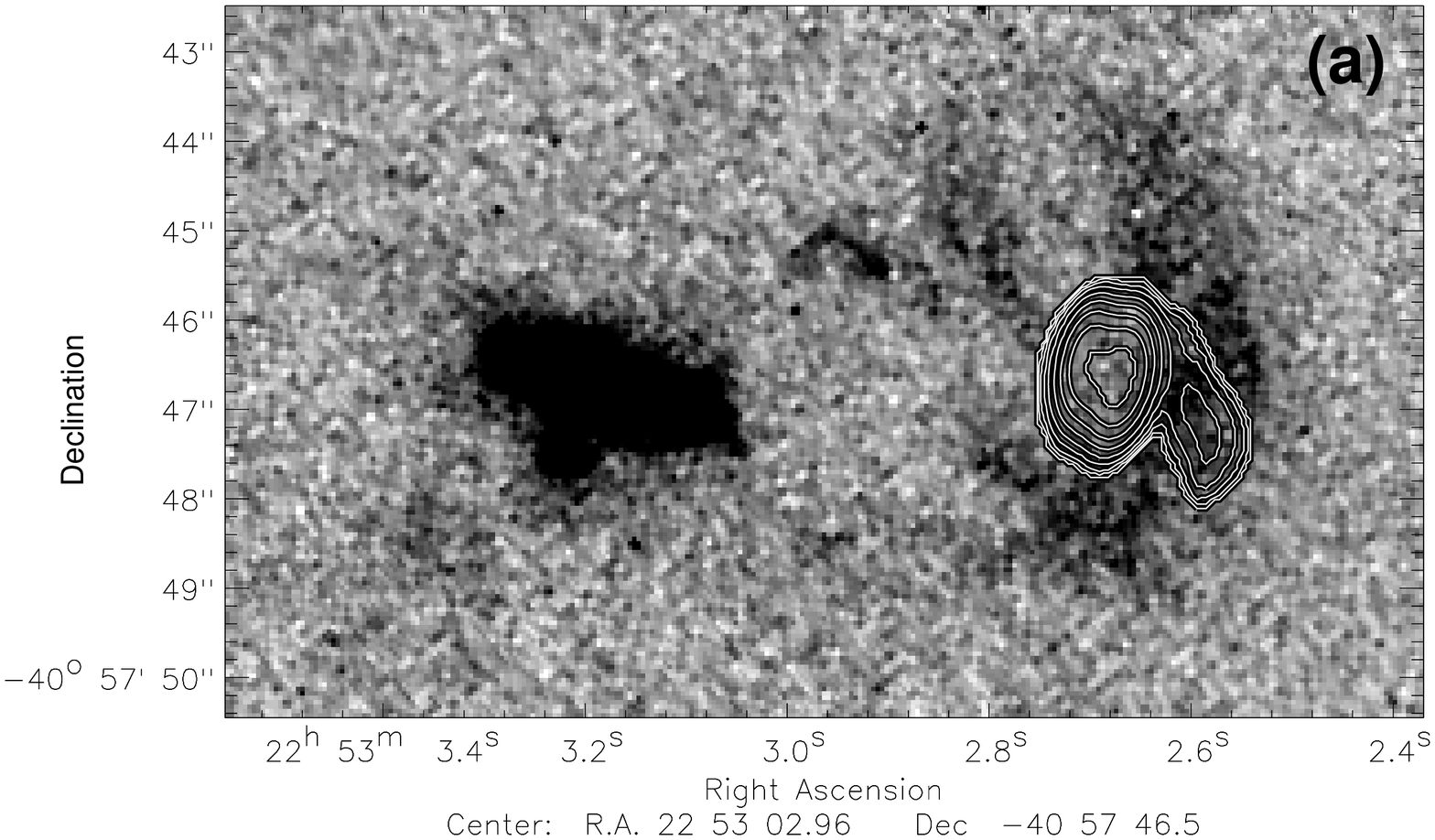}
\includegraphics{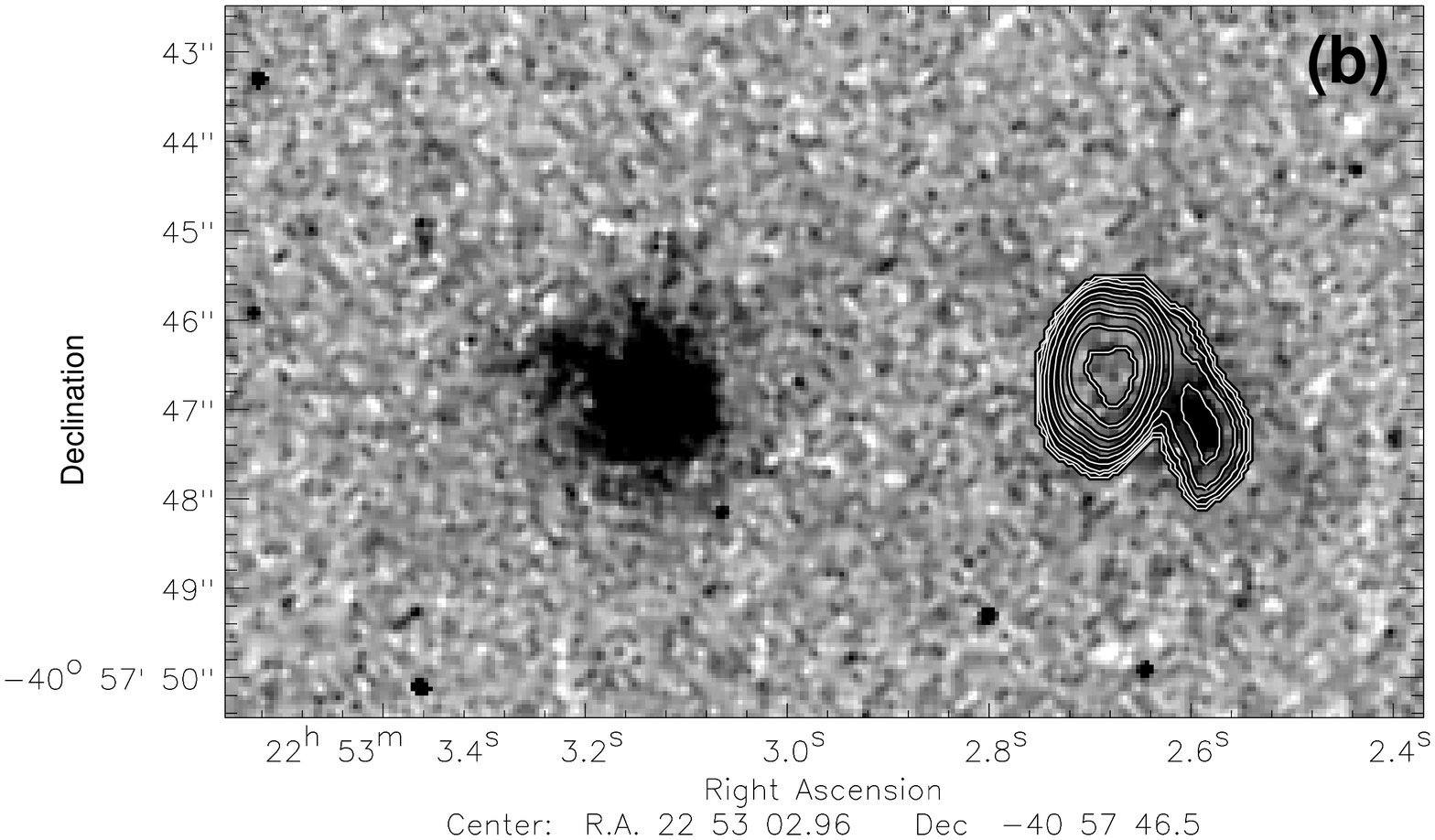}
\includegraphics{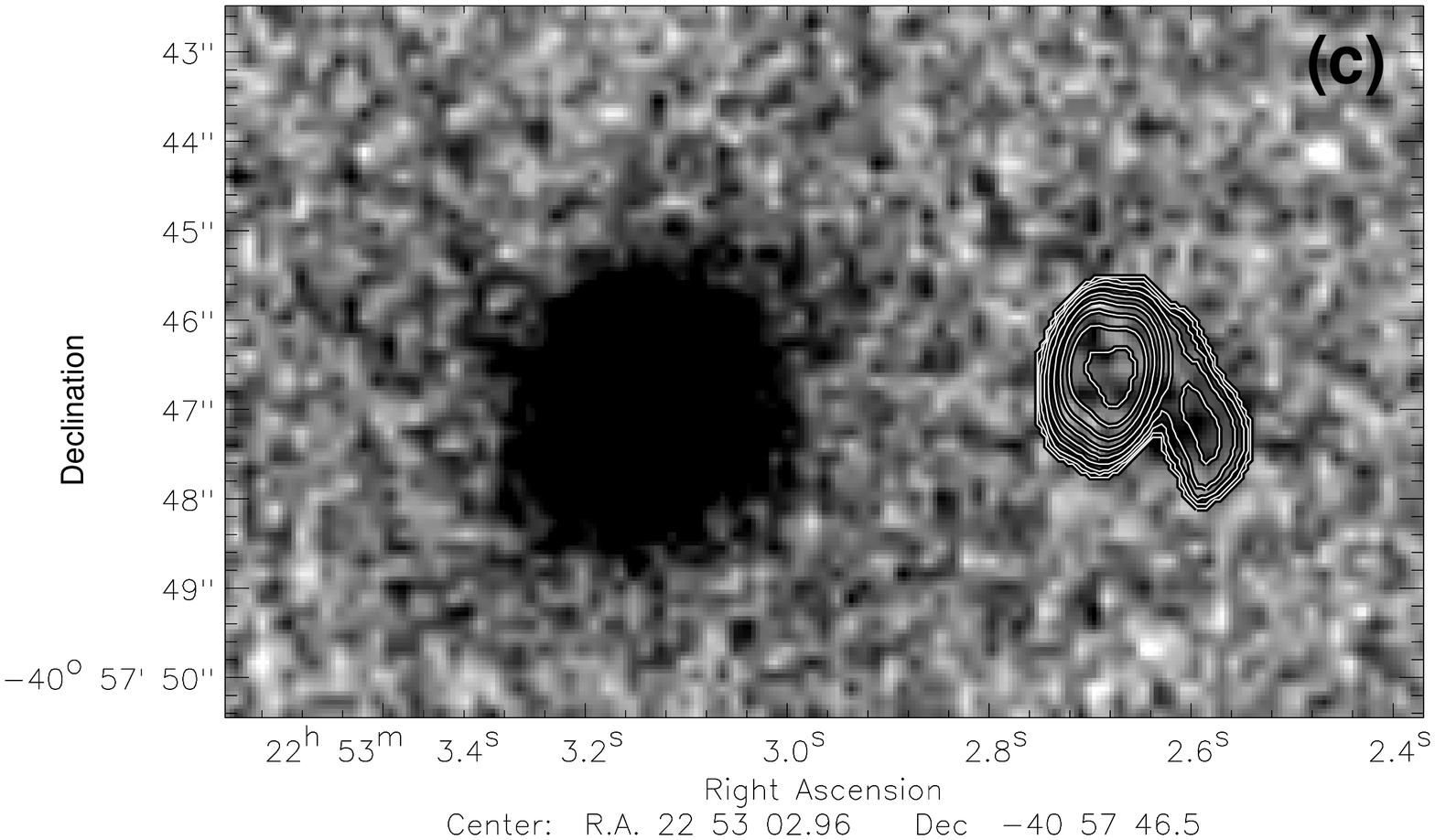}
\end{center}
\caption{(a - top) WFPC2 [O\textsc{iii}]5007\AA\ emission line image of
  PKS2250-41 overlaid with 15GHz VLA radio contours. (b - centre)
  WFPC2 F547M continuum image of PKS2250-41 overlaid with 15GHz VLA
  radio contours. (c - bottom) SOFI Ks-band image of PKS2250-41 overlaid with 15GHz VLA
  radio contours. Both the optical and infrared images display flux at the
  position of peak emission line intensity in the western arc, which
  is also coincident with the secondary hotspot within the western
  radio lobe.  The WFPC2 and radio data are taken from T05.
\label{overlays}}
\end{figure}

We have also carried out aperture photometry of the host galaxy using
the F547M WFPC2 data of T05, obtaining a host galaxy magnitude in this filter
of $20.54 \pm 0.12$.  Comparison of the relative flux densities observed
in each of the two continuum filters (F547M and $K_S$) confirms that
they are consistent with the spectral 
energy distribution expected for a 9.5Gyr old stellar population.

It is interesting to note that the filamentary line emission displayed
in Fig.~\ref{oiii_closeup} is matched by an extension in the optical
continuum observed from this galaxy along the same position angle
(Fig.~\ref{overlays}). The extension to the host galaxy is clearly
visible in the optical, and it is also noteworthy that the lowest surface brightness
IR emission is somewhat disturbed in this region of the
galaxy, perhaps reflecting a distortion in the underlying galaxy structure.

\subsection{Companion objects}

\begin{figure}
\vspace{1.9 in}
\begin{center}
\includegraphics{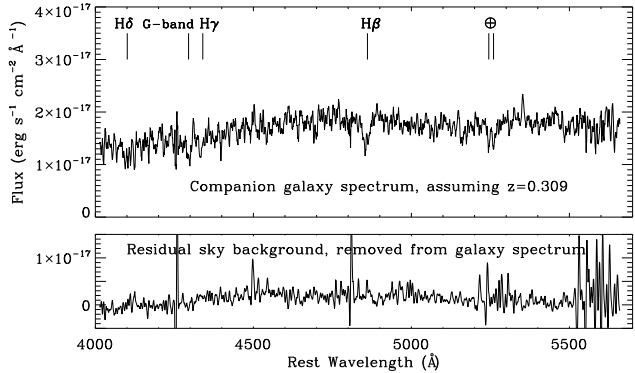}
\end{center}
\caption{Smoothed FORS1 spectrum of the companion object lying $\sim$10.5 arcsec to
  the north west of PKS2250-41, extracted from a
  $1^{\prime\prime}$ aperture centred on the continuum
  centroid. Absorption line features consistent with a
  redshift of $z=0.309$ are labelled.  We also plot the residual sky
  background.
\label{companion_spec}}
\end{figure}

Figure~\ref{companion_spec} displays the FORS1 spectrum for the
companion galaxy to the north east, extracted within a
$1^{\prime\prime}$ aperture.  The strongest feature in the spectrum is
an absorption line consistent with H$\beta$ at a
redshift of $z = 0.309$.   The continuum shape
is also generally consistent with a galaxy at the assumed redshift, and we
also see a noisy absorption features at wavelengths appropriate to
the G-band, H$\gamma$ and H$\delta$.  

Using the WFPC2 continuum data of T05 and the IR data
presented in this paper, we have carried out aperture photometry of
this source.  In the optical, this companion galaxy has a magnitude of
$20.57 \pm 0.12$, slightly fainter than the radio source host
galaxy. In the IR, the observed $K_S-$band 
magnitude is $16.48 \pm 0.08$.  The relative flux densities of the
observed-frame optical and infrared emission from this galaxy are
consistent with a single old stellar population (OSP) aged between 2-4Gyr. However, given the
presence of Balmer absorption features in this galaxy a composite
stellar population including both old and young stars is likely to
provide a more accurate description of this galaxy.

Past studies of PKS2250-41 have suggested that the prominent emission
line arc to the west of the host galaxy originates from a direct collision
between the radio source jet and a companion galaxy (Clark et al 1997,
VM99). Figure~\ref{overlays} displays the infrared and optical
continuum and the [O\textsc{iii}] emission line
imaging data for PKS2250-41 overlaid with radio contours (all data
except infrared from T05).  Continuum emission is detected in both
the $K_S$ and F547M filters within the arc, coincident with the
secondary radio hotspot.  We have determined the
flux density of this feature in both filters; the resulting ratio of 
F547M/$K_S$ emission is $24.3 \pm 58\%$.  This value is consistent
with unreddened stellar populations aged between either 0.006-0.009Gyr or
0.05-0.1Gyr, or alternatively reddened YSPs of even younger ages.

\section{Discussion}

\subsection{The evidence for jet-induced star formation.}

For many years now it has been known that radio source jets have the
potential to substantially affect the ambient interstellar/intergalactic
medium.  Separating the influence of the jet from other forms of the
alignment effect (Chambers, Miley \& van Breugel 1987, McCarthy et al
1987), i.e. scattered AGN emission or line and nebular continuum
emission associated with AGN-ionized clouds, is not always
straightforward (e.g. Tadhunter et al 2002; Inskip et al 2005) but in many cases there is
clear-cut evidence for so-called `jet--cloud interactions'.  This may
take the form of shocks associated with interactions between the radio
jet and cool gas clouds in the ISM/IGM, or alternatively, jet-induced
star formation. 

Jet-induced star formation was first proposed in order to explain the
observations of such sources as Centaurus A (Blanco et al
1975; Graham \& Price 1981; Brodie, K\"{o}nigl \& Bowyer 1983;
Sutherland, Bicknell \& Dopita 1993; Schiminovich et al 1994; Oosterloo \& Morganti
2005) and Minkowski's Object (van Breugel et al
1985; Croft et al 2006).  
Current theoretical models (e.g. Fragile et al 2004; Mellema,
Kurk \& R\"{o}ttgering 2002, and references therein) suggest that
radio source shocks propagating through the clumpy ISM/IGM trigger the
collapse and/or fragmentation of overdense regions, which may then
subsequently form stars.  
Good evidence for jet-induced star formation has also been observed in
the case of the radio sources 4C 41.17 (Dey et al
1997; Bicknell et al 2000), 3C34 (Best, Longair \& R\"{o}ttgering 1997) and 3C48 (Stockton
et al 2007), but on the whole, while evidence for shocks associated
with the radio source is relatively common, jet-induced star formation
is far less frequently observed.
PKS2250-41 is an archetypal example of a galaxy displaying jet-cloud
interactions, with clear-cut evidence for shocks associated with the
expanding radio source; specifically, via the observed variation in
ionization state and gas kinematics in the vicinity of the western
radio lobe hotspots (Clark et al 1997; VM99; T05).   However, the
origin of the emitting material 
and the case for jet-induced star formation is less absolute, and we
therefore reconsider the issue here.   

The primary evidence is the continuum emission which, in addition to
the well-studied line emission, is also observed within the western
radio lobe.  The continuum emission is approximately co-spatial with
the secondary radio source hotspot (Fig.~\ref{Fig: radio}), has low polarization (Shaw et al
1995) and only a limited contribution of nebular continuum emission
(Dickson et al 1995).  Clark et al (1997) suggested that the residual
continuum emission could originate from a late-type spiral or
irregular galaxy, with which the radio jet has collided; radio source
shocks driven through the gas clouds associated with such an object
can also account for the impressive scale and luminosity of the
observed emission line arc, and possibly also the shortness of the
western lobe relative to the eastern lobe. 

Our infrared
observations of PKS2250-41 add further weight to this scenario; the
faint detection in the western lobe lies at an identical position to
the optical continuum position.  However, the inferred spectral shape
implies that a {\it very} young stellar population is dominating the optical
emission, suggesting that the proximity of the radio source may very
well have triggered recent star formation within this object.

\subsection{Interactions in a group environment.}

Our spectroscopic observations of the companion galaxy to the north
east confirm that it lies at an almost identical redshift to
PKS2250-41 ($z = 0.3090 \pm 0.0005$ cf. $z = 0.3074$).  The companion galaxy
colours and the presence of Balmer absorption features in its spectra
suggest that it is currently forming stars.  A further interesting
feature is that the position angle of the major axis of this galaxy is
closely aligned with the line emission extending outwards from the
radio galaxy (see Fig.~\ref{oiii_closeup}).  It therefore seems highly
plausible that both galaxies form part of an interacting group,
a similar environment to that of the radio source PKS1932-46 (Inskip
et al 2007).

\subsubsection{Up close and personal: a recent interaction?}

\begin{figure}
\vspace{4.5 in}
\begin{center}
\includegraphics{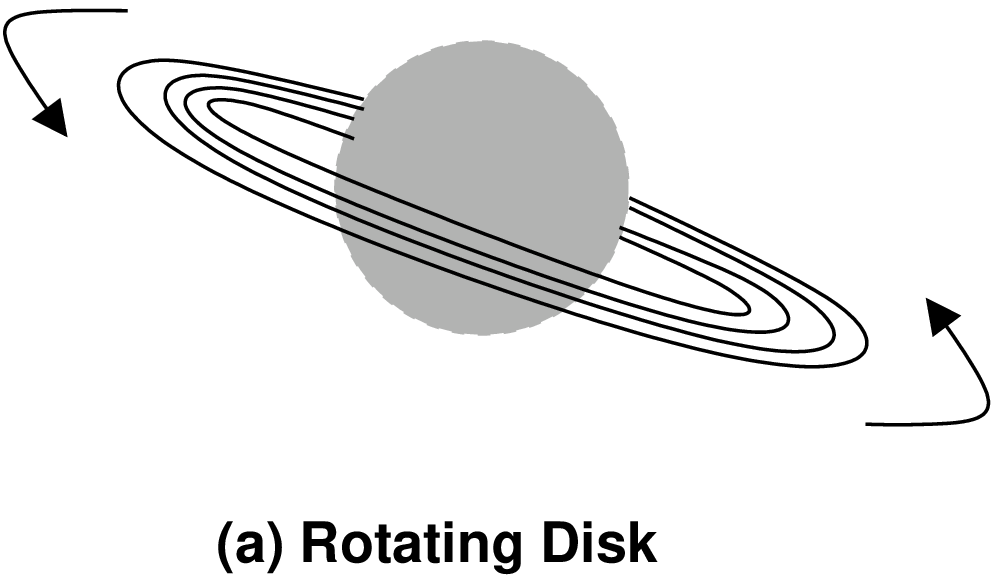}
\includegraphics{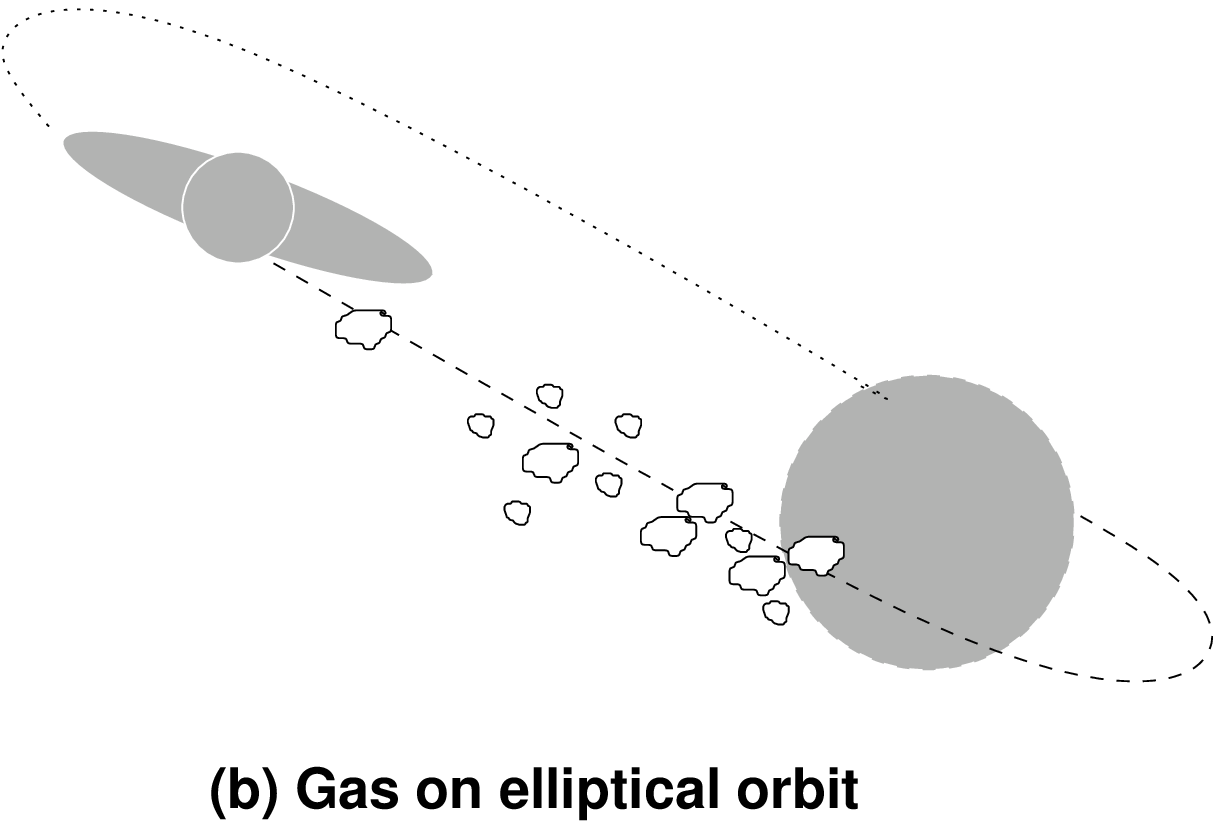}
\includegraphics{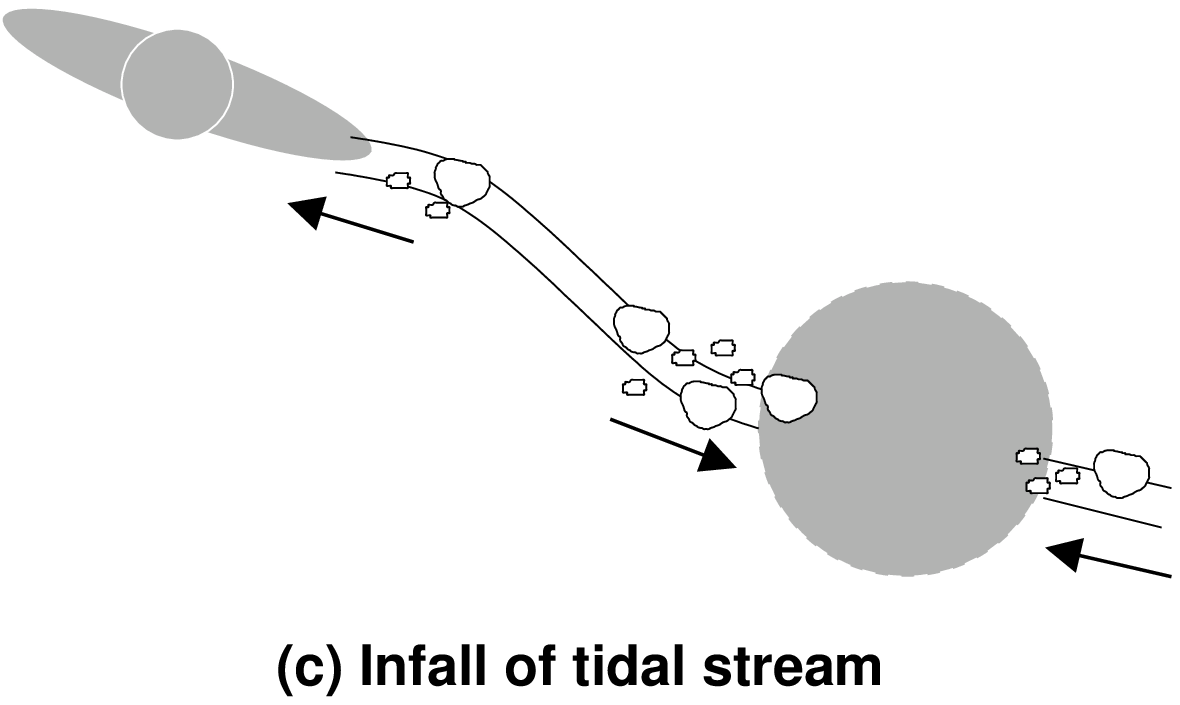}
\end{center}
\caption{Three sketches illustrating the different interpretations of
  the origin of the emission line and continuum spur (see Fig.~\ref{oiii_closeup}
  and Fig.~\ref{overlays}a,b), as described in full in section
  4.2.1. Frame (a) illustrates the case of gas in a rotating disk
  structure orbiting the host galaxy, frame (b) gas strewn along the
  elliptical orbit of the companion galaxy to the northeast, and frame
  (c) the infall of material along the tidal stream between the two
  galaxies. 
\label{cartoons}}
\end{figure}

In their paper on the emission line and continuum imaging of
PKS2250-41, Tilak et al (2005) highlighted the presence of a continuum
emission spur extending from the host galaxy, aligned towards the
neighbouring companion galaxy to the north east (see also
Fig.~\ref{overlays}b), and coincident with the brightest line emission
(see also Fig.~\ref{oiii_closeup}).   Additionally, the
position angles of both spur and companion differ by no more than
$6^{\circ}$, while the filamentary line emission surrounding the spur
is even better aligned with the companion galaxy.

The presence of this filamentary feature has profound implications for our
understanding of how this particular radio source was triggered.  At
present, there are several different mechanisms by which the fuelling
of the AGN may occur.  Fuel may be
supplied by the infall of cooling gas (Bremer, Fabian \& Crawford
1997), though this is most likely to cause a renewal of the AGN
activity in a galaxy rather than being related to the earliest
triggering of the source (Nipoti and Binney 2005). Alternatively, a
merger between two galaxies, one of which is gas-rich, can 
lead to fuelling of the AGN (e.g. Heckman et al 1986; Barnes \&
Hernquist 1996; Mihos \& Hernquist 1996); the ensuing AGN activity
may occur contiguously (e.g. di Matteo, Springel \& Hernquist 2005) or
after a substantial delay (e.g. Sanders et al 1988), and in many
cases is associated with the presence of young- to intermediate-aged
stellar populations (e.g. Tadhunter et al 2005). In the case of
PKS2250-41, we may indeed be observing 
direct physical evidence of the specific event which led to the
triggering of the radio source activity within the galaxy.

While our data suggest that the observed filamentary feature could be consistent with a
large-scale rotating structure (e.g Fig.~\ref{cartoons}a), there is an order of magnitude
discrepancy between the stellar mass estimated from the K-band
magnitude and the mass estimated from the `rotation curve' of the
emission lines.  If the observed feature is indeed a rotating disk,
the morphology of the emission would require it to be nearly egde-on.
Given this fact and the mass discrepancy, we thus rule out a fully
sampled rotating disk in dynamical equilibrium with the host potential
being the source of the highest surface brightness aligned features.
However, it is possible that we might be observing a disk feature
which is not fully sampled, and may only be observing the part of it
which lies within the ionization cone of the quasar.

The observations might instead be better explained by gas strewn along the
elliptical orbit of the companion, which is seen almost edge-on (see Fig.~\ref{cartoons}b).    In
this case, the gas observed close to the nucleus of PKS2250-41 could
be moving almost perpendicularly to the line of sight, particularly if
this gas doesn't sample the perigee of the companion's orbit. The
higher surface brightness of the emission closer to the 
nucleus would be due to the gas being lit up by the quasar; the offset
of the emission from the radio axis is naturally explained in this
scenario.  The
exact form of the velocity curve would depend on the distrubution of
gas around the orbit, and could easily account for the observations. 

Another option is a tidal streamer pulled off from the disk of the
radio source precursor object by the companion galaxy (see, e.g., the
numerical simulations of di Matteo et al (2005)). At some stages
following the first pass of the two merging galaxies the gas will
stream back (almost radially) towards the nuclei of the galaxies (see Fig.~\ref{cartoons}c).  In
this case, the velocity curve could reflect the streaming infall
motions on either side of the nucleus.  Such infalling material could
be responsible for triggering the AGN/jet activity within PKS2250-41. 

\subsubsection{The triggering of the activity}

Spitzer MIPS photometry of a complete sample of 2Jy sources (Tadhunter
et al 2007) has been used to address the issue of the dominant dust
heating mechanism in AGN.  The [O\textsc{iii}]5007\AA\ emission line
luminosity (which can be used as a tracer of AGN power, e.g. Rawlings
\& Saunders 1991; Tadhunter et al 1998; Simpson 1998) is strongly
correlated with the mid-IR emission (see Fig. 1 in Tadhunter et al 2007),
indicating that AGN illumination is the principal dust-heating
mechanism.  The mid- to far-IR fluxes of PKS2250-41 (22mJy and
11.6 mJy at 70$\mu m$ and 24$\mu m$ respectively) imply an overall far-infrared luminosity of $L_{IR}
\sim 2\times 10^{11}L_{\odot}$ (Sanders \& Mirabel 1996).  Although
this is consistent with a LIRG-class source, this source falls firmly on the correlation
between mid- to far-IR emission and the [O\textsc{iii}]5007\AA\ emission line
luminosity, suggesting that in the case of this source the   mid- to
far-IR luminosity is derived from AGN rather than starburst
illumination.  Further to this, the mid- to far-IR emission is well centered on the centroid of the radio source
host galaxy, with no displacement towards the companion such as might
be expected if the companion galaxy was actively forming stars at a
high rate.  

It is clear from these results that PKS2250-41 is not a case of a
source triggered in the final stages of a major galaxy merger.
However, the scenario wherein a luminous AGN has been triggered by an
encounter with another galaxy, some time after the point of closest
approach, is far more plausible.   Neither the radio galaxy nor its
companion can be classed as a very massive elliptical.  Additionally,
the lack of very powerful far-IR emission from either source is
consistent with low levels of starburst activity, with the AGN being
the dominant cause of dust heating.  If this is a case of triggering
following a (first pass) galaxy 
encounter, the companion galaxy would most likely
be on a highly elliptical orbit, and eventually merge with the radio
source host at a much later point in time. Many of the different
merger models predict relatively low levels of star formation until
the nuclei finally merge (e.g. Hernquist \& Mihos 1995, Springel et al 2005).
 The tidal forces
associated with the initial encounter could be expected to cause the formation
of the filamentary spur, while gas settling
onto the central black hole for some time after the time of closest
approach could trigger the AGN activity.  Similar features are
observed in the case of PKS0349-27 (Danziger et al 1984), which
displays a tidal bridge linking it to a companion galaxy.

It is also interesting to contrast the results for PKS2250-41 with those
of the z=0.23 FRII radio galaxy PKS1932-464 (Inskip et al 2007),
which, like PKS2250-41, exists within an
interacting group environment and  also displays a spectacular EELR.
However, the similarities end there.  Despite the materially enriched IGM, the
radio source of PKS1932-464 does not interact strongly with its
environment.  Additionally, while we observe little in the way of star
formation associated with PKS2250-41, evidence for significant star formation is
seen in the IGM surrounding PKS1932-464, as well as within the
neighbouring companion galaxy.  These results illustrate that even
when AGN are triggered under apparently similar circumstances, a
diversity of different outcomes are possible.

\section{Conclusions}

We have carried out a multiwavelength study of the radio source
PKS2250-41, for the first time combining imaging, long-slit and IFU spectroscopic
observations of this source.  In addition to analysing the EELR in
greater detail, we have now  also been able to constrain the nature of the
companion objects surrounding the host galaxy, as well as the nature
of the host galaxy itself. This has allowed us to produce a far
clearer picture of the different astrophysical processes ongoing within this source,
and how the activity was triggered.

The key results are as follows:

\begin{enumerate}
\item[$\bullet$]{}Our VIMOS IFU data produce results which are
  consistent with previous long-slit spectroscopy, allowing us to
  trace the different kinematic components across the EELR and
  confirming the effects of the jet-cloud interactions occuring within
  this source.  We observe lower
  ionzation level for the gas and broader lines in the vicinity of the
  western hotspots than elsewhere within the EELR, consistent with the
  presence of shocks in those regions.  The remainder of the EELR gas
  is predominantly photoionized.  In terms of the gas kinematics, we
  detect a north-south velocity gradient across the EELR and boosting
  of the line widths along the radio axis, particularly close to the
  western lobe hotspots. Narrower line emission is observed between
  the radio galaxy and the companion galaxy to the northeast.
\item[$\bullet$]{}The host galaxy is an $\sim 1.4 M_{\star}$ elliptical,
  and part of a group of galaxies, including a disk galaxy of the same
  redshift to the north-east, and a faint companion object currently
  interacting with the radio source in the vicinity of the western
  lobe secondary hotspot.
\item[$\bullet$]{}Our detection of this faint companion object in the
  infrared allows us to constrain its SED.  The recent star-formation
  implied by this result suggests that jet-induced star formation is
  indeed occuring within  this object.
\item[$\bullet$]{}The filamentary structure extending to the north
  east outwards from   the host galaxy is potentially linked with the
  triggering of the current AGN activity 
  in this source.  We believe that a tidal encounter with the
  companion object is the most likely cause of the triggering of the
  activity.
\end{enumerate}

This
object provides some of the best evidence that radio source activity
can be triggered in a tidal encounter with a companion, before the
final merger, but after the first pass of the nuclei in interacting
galaxies.  PKS2250-41 is clearly a most remarkable object, and a showcase for
many aspects of radio source physics.  But perhaps its greatest
advantage is its ability to further our understanding of both the
triggering of radio source activity, and also of the influence that a
radio source has on the material in its surrounding environment.

\section*{Acknowledgements}
KJI acknowledges support from a PPARC research fellowship and JH a
PPARC PDRA. The work of MV-M  has been supported by the Spanish
Ministerio de Educaci\'on y Ciencia and the Junta de Andaluc\'ia
through the grants AYA2004-02703 and TIC-114. The IFS data published
in this paper have been reduced using VIPGI, designed by the VIRMOS
Consortium and developed by INAF Milano; KJI would particularly like
to thank Paolo Franzetti for guidance in the use and reliability of
the VIPGI software.   We also thank the ESO technical and support
staff for indulging our request for the use of SOFI in the small field
mode.  This research has made use of the NASA/IPAC Extragalactic
Database (NED) which is operated by the Jet Propulsion Laboratory,
California Institute of Technology, under contract with the National
Aeronautics and Space Administration.   We thank the anonymous referee
for some very useful comments which greatly improved the clarity of
this paper.

\appendix
\section{Emission line fluxes from the FORS1 spectroscopy}
Here we present the emission line fluxes from the 1.6 arcsec apertures
extracted from our FORS1 long slit spectroscopic observations.

\begin{table*}
\caption{Spectroscopic properties of PKS2250-41, determined from
  spectra extracted in successive 1.6$^{\prime\prime}$ apertures across
  the slit.  
  The total integrated line fluxes are given in units of
  $10^{-17}\rm{erg\,s^{-1}cm}^{-2}$. Non-detected lines (denoted by a dash) would have a maximum
  flux of the order of $1 \times 10^{-17}\rm{erg\,s^{-1}cm}^{-2}$. Errors are dominated by flux-calibration
  errors (estimated to 
  be $\la 10$\%); the listed error values are determined by adding
  this calibration error in quadrature with the errors arising from photon
  statistics. All quoted values
  in this table are corrected for galactic extinction using the
  $E(B-V)$ for the Milky Way (Schlegel, Finkbeiner \& Davis 1998) and
  the empirical extinction function of Cardelli et al (1989).}
\begin{center} 
\begin{tabular}{ccccccccc}
Spectrum:& Redshift&[SII]& H$\delta$ & H$\gamma$ & [O\textsc{iii}]  &
HeII &
H$\beta$ & [O\textsc{iii}] \\
& & 4072\AA & 4102 \AA & 4340\AA & 4363\AA & 4686\AA & 4861\AA & 5007\AA\\
\hline
S1: $- 6.4^{\prime\prime}$ &0.3083& --& -- & -- & -- & -- & $3.97 \pm 0.96$& $39.64 \pm
4.02$\\
S2: $- 4.8^{\prime\prime}$ &0.3083& -- & $4.69 \pm 1.18$ & $9.90 \pm 1.44$ & $< 1.67$ & -- & $16.59 \pm 1.98$& $90.45 \pm
9.09$\\
S3: $- 3.2^{\prime\prime}$ & 0.3081& --& -- & $3.88 \pm 0.99$ & $< 1.94$ & -- & $7.18 \pm 1.40$& $32.85 \pm 3.42$\\
S4: $- 1.6^{\prime\prime}$ & 0.3082 & --& -- & $4.92 \pm 1.10$ & $< 2.46$ & -- & $13.01 \pm 1.80$& $96.98 \pm 10.00$\\
S5: $+ 0.0^{\prime\prime}$ &0.3076 & $4.37 \pm 0.95$& $14.36 \pm 1.98$ & $27.58
\pm 3.15$ & $15.43 \pm 2.19$ & $19.74 \pm 2.24$ & $62.29 \pm 6.34$ &$696.22 \pm 69.74$\\
S6: $+ 1.6^{\prime\prime}$ &0.3073& -- & $10.24 \pm 1.54$ & $22.41
\pm 2.52$ & $10.78 \pm 1.44$& $14.24 \pm 1.70$ & $46.80 \pm
4.80$ & $551.08 \pm 55.25$\\
S7: $+ 3.2^{\prime\prime}$ &0.3076 & --& -- & $3.62 \pm 0.92$ &-- & -- & $3.87 \pm 1.32$ & $37.14 \pm 3.84$\\
S8: $+ 4.8^{\prime\prime}$ &0.3077& --& -- & -- & -- & --& $1.33 \pm 0.64$ & $7.76 \pm 1.02$\\
S9: $+ 6.4^{\prime\prime}$ &0.3076& --& -- & -- & -- & -- & --& $3.16 \pm 0.72$\\
S10: $+ 8.0^{\prime\prime}$& 0.3074& --& -- & --& -- &-- & --& $10.06
\pm 1.25$\\
S11: $+ 9.6^{\prime\prime}$& 0.3084& -- & -- & -- &--  & --& -- & $2.25
\pm 0.76$\\

\end{tabular}                      
\end{center}                       
\end{table*}

\label{lastpage}

\end{document}